\documentclass[pdflatex,sn-mathphys-num]{sn-jnl}

\usepackage{graphicx}%
\usepackage{multirow}%
\usepackage{amsmath,amssymb,amsfonts}%
\usepackage{amsthm}%
\usepackage{mathrsfs}%
\usepackage[title]{appendix}%
\usepackage{xcolor}%
\usepackage{textcomp}%
\usepackage{manyfoot}%
\usepackage{booktabs}%
\usepackage{algorithm}%
\usepackage{algorithmicx}%
\usepackage{algpseudocode}%
\usepackage{listings}%

\usepackage{ulem}
\usepackage{bm}
\usepackage{caption}
\usepackage{subcaption}
\usepackage{mathtools}
\usepackage{url}

\def\finalversion{} 

\ifdefined\finalversion
  \newcommand{\added}[1]{#1}
  \newcommand{\deleted}[1]{}
  
\else
  \newcommand{\added}[1]{{\color{red}#1}}
  \newcommand{\deleted}[1]{{\color{red}\sout{#1}}}
  
\fi

\theoremstyle{thmstyleone}%
\theoremstyle{thmstyletwo}%

\theoremstyle{thmstylethree}%

\begin{document}

\title[The effect of the number of parameters and local feature patches on loss landscapes in distributed QNNs]{The effect of the number of parameters and the number of local feature patches on loss landscapes in distributed quantum neural networks}


\author*[1]{\fnm{Yoshiaki} \sur{Kawase}}\email{ykawase@g.ecc.u-tokyo.ac.jp}

\affil*[1]{\orgdiv{Graduate School of Information Science and Technology}, \orgname{The University of Tokyo}, \orgaddress{\street{7-3-1 Hongo, Bunkyo-ku}, \city{Tokyo}, \postcode{113-8656}, \country{Japan}}}


\abstract{
Quantum neural networks hold promise for tackling computationally challenging tasks that are intractable for classical computers. 
However, their practical application is hindered by significant optimization challenges, arising from complex loss landscapes characterized by barren plateaus and numerous local minima. 
These problems become more severe as the number of parameters or qubits increases, hampering effective training. 
To mitigate these optimization challenges, particularly for quantum machine learning applied to classical data, 
we employ an approach of distributing overlapping local patches across multiple quantum neural networks, processing each patch with an independent quantum neural network, and aggregating their outputs for prediction. 
In this study, we investigate how the number of parameters and patches affects the loss landscape geometry of this distributed quantum neural network architecture via \added{theoretical and empirical} Hessian \added{analyses} \deleted{analysis} and loss landscape visualization. 
Our results confirm that increasing the number of parameters tends to lead to deeper and sharper loss landscapes. 
Crucially, we \added{theoretically derive and empirically} demonstrate that increasing the number of patches significantly reduces the largest Hessian eigenvalue at minima. 
\added{Furthermore, our analysis of the full Hessian eigenspectrum reveals a structure consisting of a bulk of near-zero eigenvalues and distinct outlier spikes corresponding to the number of classes, similar to classical deep learning models.}
\added{These findings suggest} \deleted{This finding suggests} that our distributed patch approach acts as a form of implicit \added{structural} regularization, promoting optimization stability and potentially enhancing generalization. 
Our study provides valuable \added{theoretical and empirical} insights into optimization challenges and highlights that the distributed patch approach is a promising strategy for developing more trainable and \added{scalable} \deleted{practical} quantum machine learning models for classical data tasks. 
}




\maketitle

\section{Introduction}\label{sec:intro}
Variational quantum algorithms~\cite{cerezo2021variational} are widely studied as promising applications of near-term quantum computing.  
Notable examples include variational quantum eigensolvers~\cite{kandala2017hardware, peruzzo2014variational}, 
quantum approximate optimization algorithms~\cite{farhi2014quantum}, and quantum neural networks (QNNs)~\cite{farhi2018classification, mitarai2018quantum}. 
This work focuses specifically on QNNs, which possess high expressibility and potentially enable solving challenging tasks for classical machine learning~\cite{abbas2021power, holmes2022connecting}. 
However, QNN training is hindered by critical challenges, including the prevalence of numerous local minima~\cite{you2021exponentially, bittel2021training} and the barren plateau phenomenon,  
where gradients vanish exponentially with system size~\cite{mcclean2018barren}, exacerbated by cost function design~\cite{cerezo2021cost} and hardware noise~\cite{wang2021noise}. 

To understand and mitigate these optimization challenges, characterizing the geometry of the loss landscape is essential.
Techniques adapted from classical deep learning, such as loss landscape visualization and Hessian analysis, are valuable for this purpose. 
Visualization offers qualitative insights into training dynamics influenced by architecture or optimizers~\cite{li2018visualizing, zhao2020bridging, garipov2018loss, fort2019deep}; 
Hessian analysis, which quantifies local curvature and the sharpness of minima, is particularly useful for evaluating the influence of specific network components~\cite{yao2020pyhessian} or assessing robustness to parameter perturbations~\cite{zhao2020bridging}. 
These tools have already been applied in quantum machine learning to study the effects of ansatz structure, data encoding, and initialization~\cite{rudolph2021orqviz, huembeli2021characterizing}.

One proposed strategy to alleviate training difficulties involves distributed QNNs~\cite{pira2023invitation}, using multiple smaller QNNs. 
For instance, a circuit cutting technique~\cite{bravyi2016trading, peng2020simulating} allows us to simulate or execute large circuits using smaller ones, 
but they incur an overhead in the number of small circuit evaluations that grows exponentially with the number of cutting gates. 
While mitigation techniques for the exponential cost are being explored, such as approximations for machine learning tasks~\cite{marshall2023high}, 
this cost remains a significant scalability challenge. 
To circumvent this overhead, particularly for classical data, 
our previous work introduced an alternative approach~\cite{kawase2024distributed}: 
partitioning input features into local patches (subregions of data) and distributing these patches across multiple independent QNNs, 
subsequently aggregating their outputs for classification. 
This method, which avoids circuit cutting overhead, achieved high accuracy on $10$-class classification on the MNIST dataset. 

In this paper, we analyze the loss landscape geometry of this distributed patch-based QNN model, 
specifically investigating how the number of variational parameters and the number of patches affect the loss landscape characteristics relevant to optimization. 
Building on our previous work~\cite{kawase2024distributed}, 
we distribute overlapping input patches to individual QNNs, aggregate their outputs, and apply the Softmax function for classification. 
Through loss landscape visualization and \added{theoretical and empirical} Hessian \added{analyses} \deleted{analysis}, 
we find that increasing the number of variational parameters leads to deeper and sharper loss landscapes around minima. 
Notably, we \added{theoretically derive and empirically} demonstrate that increasing the number of patches reduces the largest Hessian eigenvalue evaluated at the minima.
\added{Specifically, our theoretical analysis reveals that the Hessian matrix can be decomposed into two terms scaling down by factors of $1/n_\text{qc}^2$ and $1/n_\text{qc}$ (where $n_\text{qc}$ is the number of patches). 
This mathematically proves that, unlike single global QNNs suffering from vanishing gradients or sharpened minima, 
this distributed architecture provides implicit structural regularization. 
Furthermore, we statistically validate these findings across multiple random seeds and investigate the full Hessian eigenspectrum. 
We observe a bulk with near-zero eigenvalues and distinct outlier spikes corresponding to the number of classes, similar to classical deep learning models. }
\added{These} \deleted{This} finding\added{s} suggest\deleted{s} that our distributed overlapping local patch approach mitigates the sharpness around minima, potentially enhancing trainability and generalization\deleted{, acting as a form of implicit regularization}. 
We anticipate that these findings offer valuable insights into optimization challenges in quantum machine learning, contributing to the advancement of \added{scalable} \deleted{practical} quantum machine learning for classical data.

\section{Methods}\label{sec:methods}
\subsection{Distributed QNNs with Local Patches}\label{methods:distqnns}
This subsection details the $n_\text{class}$-class classification model employed in our numerical experiments, 
based on the distributed QNN architecture from our prior work~\cite{kawase2024distributed}. 
The central concept is to partition an input image into overlapping local patches and process each patch with an independent QNN to effectively capture local features across the image.

First, we describe the details of the input processing. 
We use $M \times M$ grayscale images. 
Let $x'_{i,h,w}$ represent the pixel value at row $h$ and column $w$ ($0 \le h, w < M$) of the $i$-th input image.
To enable the model to distinguish patches based on their location, similar to position embeddings in transformers~\cite{dufter2022position}, 
we incorporate positional information by adding a unique learnable bias term $b'_{h,w}$ to each pixel value $x'_{i,h,w}$ corresponding to its grid coordinates $(h,w)$.

\begin{figure}
     \centering
     \includegraphics[width=0.25\textwidth]{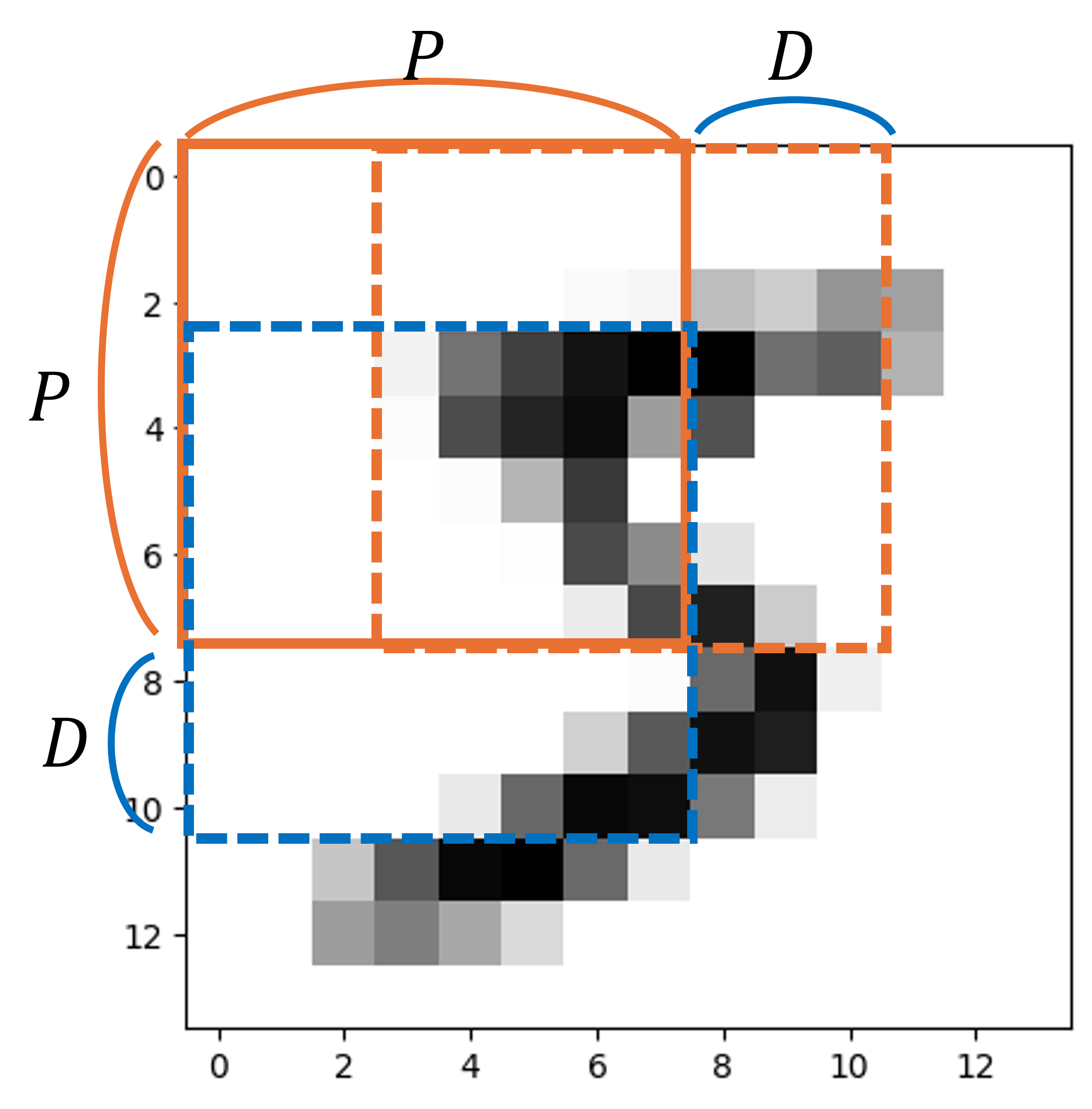}
     \caption{
     This figure illustrates the process of obtaining patches from an image. 
     First, we extract a $P\times P$ patch from the top-left corner. 
     Then, we shift the window $D$ pixels to the right to extract the next patch.  
     This horizontal extraction is repeated across the row until $L$ patches are obtained.  
     After completing one row, we shift the starting position $D$ pixels downward. 
     We then repeat the same horizontal extraction process for this new row. 
     This procedure continues row by row until a total of $L^2$ patches have been extracted.
     }
     \label{fig:patch}
\end{figure}

Next, to extract local features, we partition each image into $P\times P$ patches with a stride $D$ (see Fig.~\ref{fig:patch}). 
This process yields a total of $n_\text{qc}=L^2$ overlapping patches per image, 
where $L=(M-P)/D+1$ is the number of patches along one dimension. 
Each patch is indexed by $p \in\{0,…,n_\text{qc}-1\}$, and $n_\text{qc}$ is also the number of independent QNNs. 
The $p$-th patch covers the image region starting at top-left grid indices $(Dk,Dj)$, where $p = kL+j$, for $k,j=0,\ldots,L-1$. 
For the $p$-th patch of the $i$-th image, the $P^2$ pixel values within this patch are then flattened into a feature vector.
After adding the corresponding bias terms, 
this results in the feature vector $\bm{x}_{i,p} \in \mathbb{R}^{P^2}$:
\begin{align} \label{eqn:patch}
\bm{x}_{i,p}=( x^\prime_{i, Dk, Dj}+b^\prime_{Dk, Dj}, &\ldots,x ^\prime_{i, Dk, Dj+P-1}+b^\prime_{Dk, Dj+P-1}, \nonumber \\
&\dots \nonumber \\
x^\prime_{i, Dk+P-1, Dj}+b^\prime_{Dk+P-1, Dj}, &\ldots, x^\prime_{i, Dk+P-1, Dj+P-1}+b^\prime_{Dk+P-1, Dj+P-1}).
\end{align}
This vector $\bm{x}_{i,p}$ serves as the input to the $p$-th QNN. 

Each of the $n_\text{qc}$ QNNs operates independently. 
The $p$-th QNN employs $n$ qubits and is parameterized by its own set of variational parameters $\bm{\Phi}_p$. 
It takes the feature vector $\bm{x}_{i,p}$ as input,  
encodes it into a quantum state via angles of parameterized gates, 
and evolves the state using parameterized unitary gates. 
It computes a vector of $n_\text{class}$ expectation values $\bm{y}_{i,p} \in \mathbb{R}^{n_\text{class}}$ for a set of observables $\{O_k\}_{k=0}^{n_\text{class}-1}$:
\begin{align} 
    \bm{y}_{i,p}=\begin{pmatrix}
    \langle 0^{\otimes n}| U^\dagger (\bm{x}_{i,p},\bm{\Phi}_p)  O_0 U(\bm{x}_{i,p},\bm{\Phi}_p) |0^{\otimes n}\rangle \\
    \vdots \nonumber \\
    \langle 0^{\otimes n}| U^\dagger (\bm{x}_{i,p},\bm{\Phi}_p)  O_{ {n_\text{class}-1} } U(\bm{x}_{i,p},\bm{\Phi}_p) |0^{\otimes n}\rangle \end{pmatrix}
\end{align}

Subsequently, these output vectors $\{ \bm{y}_{i,p} \}_{p=0}^{n_\text{qc}-1}$ from all $n_\text{qc}$ QNNs are then averaged to produce a single representative vector for the $i$-th image:
\begin{align}\label{eqn:averaged_outputs} %
\overline{\bm{y}}_i = \frac{1}{n_\text{qc}} \sum_{p=0}^{n_\text{qc}-1} \bm{y}_{i,p}.
\end{align}
Finally, this averaged vector $\overline{\bm{y}}_i$ is scaled by a fixed constant value $c$ and then passed through the Softmax function to obtain the final probability distribution $\text{Prob}_i$ over the $n_\text{class}$ classes for the $i$-th input image:
\begin{align}\label{eqn:class_prob}
\text{Prob}_i = \text{Softmax}(c \cdot \overline{\bm{y}}_i).
\end{align}
The scaling by a fixed constant $c$ serves to adjust the input range of the Softmax function. 

This distributed patch-based QNN architecture provides the framework for our investigation into how the number of parameters and patches influence the loss landscape.

\begin{figure}
     \centering
     \includegraphics[width=\textwidth]{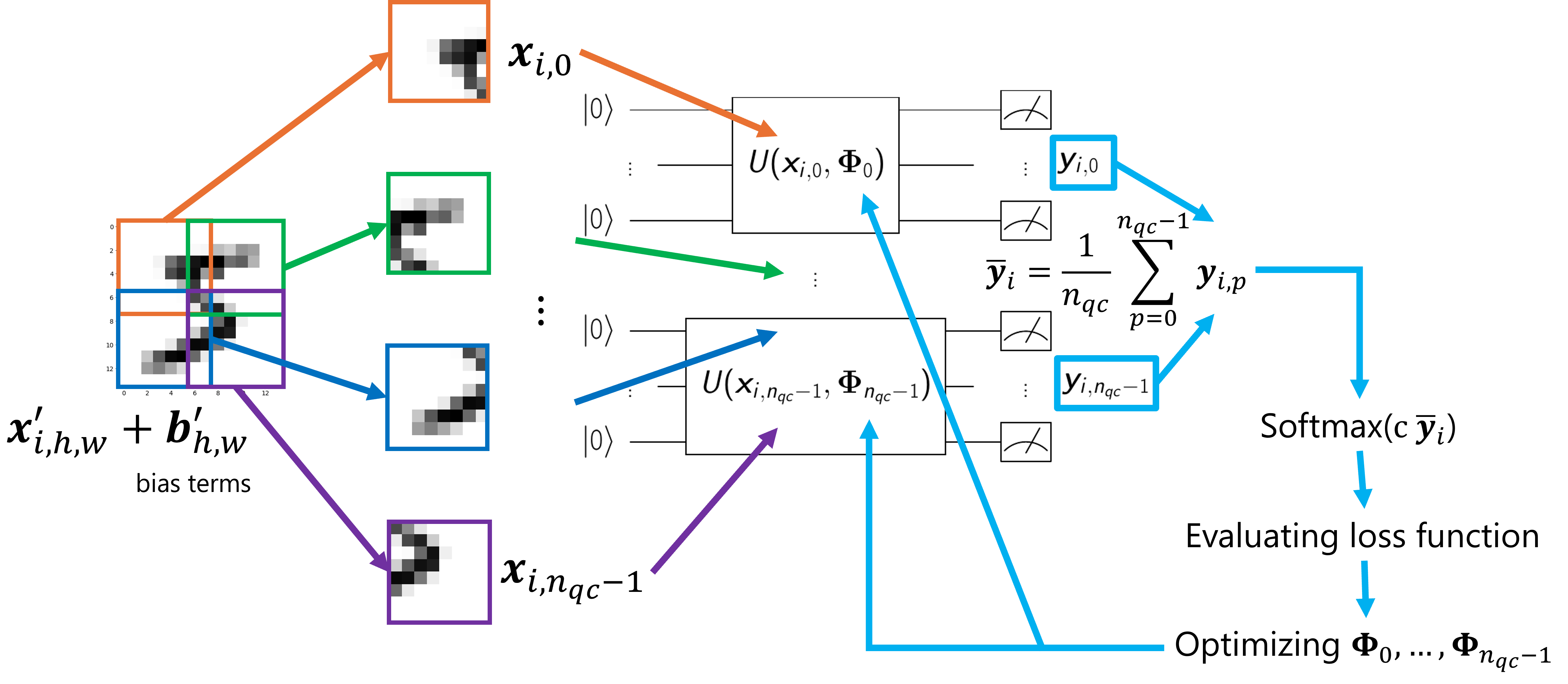}
     \caption{
     Overview of the model used in our numerical experiments: 
     First, an input image is partitioned into smaller patches. 
     Each patch is added with positional bias terms $b'_{h,w}$, 
     and processed by one of $n_\text{qc}$ different QNNs, parameterized by $\bm{\Phi}_p$. 
     Then, the expectation values from the QNNs are averaged. 
     This averaged vector $\overline{\bm{y}}_i$ is scaled and passed through the Softmax function to obtain classification predictions. 
     Finally, using these predictions, the cross-entropy loss function is evaluated, and all trainable parameters, including parameters $\bm{\Phi}_0,\ldots,\bm{\Phi}_{n_\text{qc}-1}$ and bias terms $\bm{b}'_{h,w}$, are optimized to minimize this loss. 
     This figure is adapted from Ref.~\cite{kawase2024distributed}, CC BY 4.0. }
     \label{fig:model_overview}
\end{figure}

\subsection{\added{Theoretical Hessian Analysis of Distributed QNNs}}\label{methods:theory}
\added{
In this subsection, we provide a theoretical Hessian analysis of the distributed QNN model introduced in Section~\ref{methods:distqnns}. 
Let $C$ denote the loss function. 
Using the chain rule, the gradient of the loss function $C$ with respect to a parameter $\theta_k$ is given by:
\begin{equation}
    \frac{\partial C}{\partial \theta_k} = \frac{1}{n_\text{qc}} 
    \left( \frac{\partial C}{\partial \bar{\bm{y}}_i^\top} \right) 
    \left( \sum_{p=0}^{n_\text{qc}-1} \frac{\partial \bm{y}_{i,p}}{\partial \theta_k} \right).
    \label{eqn:general_grad}
\end{equation}
By further differentiating this gradient with respect to a parameter $\theta_j$, 
we obtain the Hessian matrix elements: 
\begin{equation}
    \frac{\partial^2 C}{\partial \theta_j \partial \theta_k} = 
    \frac{1}{n_\text{qc}^2} 
    \left( \sum_{p=0}^{n_\text{qc}-1} \frac{\partial \bm{y}_{i,p}^\top}{\partial \theta_j} \right) 
    \left(\frac{\partial^2 C}{\partial \bar{\bm{y}}_i \partial \bar{\bm{y}}_i^\top} \right)
    \left( \sum_{q=0}^{n_\text{qc}-1} \frac{\partial \bm{y}_{i,q}}{\partial \theta_k} \right) 
    + \frac{1}{n_\text{qc}} \left( \frac{\partial C}{\partial \bar{\bm{y}}_i^\top} \right) 
    \left( \sum_{p=0}^{n_\text{qc}-1} \frac{\partial^2 \bm{y}_{i,p}}{\partial \theta_j \partial \theta_k} \right).
    \label{eqn:general_hessian}
\end{equation}
}
\added{
Eq.~\eqref{eqn:general_hessian} highlights the difference between using our distributed independent QNNs versus employing a single global QNN or sharing parameters across multiple smaller QNNs. 
In a standard single global QNN, which corresponds to assigning $n_\text{qc}=1$, the scaling factors $1/n_\text{qc}$ and $1/n_\text{qc}^2$ simply become $1$, offering no flattening effect. 
Alternatively, if we were to use a single global QNN that computes and aggregates numerous expectation values ($n_\text{qc} > 1$), 
the summations $\sum_{p}$ and $\sum_{q}$ in Eq.~\eqref{eqn:general_hessian} would accumulate many non-zero terms. 
These accumulated terms would largely cancel out the scaling factors $1/n_\text{qc}$ and $1/n_\text{qc}^2$, making the loss landscape more complex. 
In addition, a single global QNN naturally requires a large number of qubits, 
leading to vanishing gradients \cite{mcclean2018barren} despite the increased expressibility. 
Thus, both the complex loss landscape and the vanishing gradients make the training more difficult.
}

\added{
Similarly, if we were to share parameters across multiple smaller QNNs, 
the output $\bm{y}_{i,p}$ for multiple patches would depend on the same parameter $\theta_k$. 
Consequently, the derivatives $\frac{\partial \bm{y}_{i,p}}{\partial \theta_k}$ would be non-zero for many patches $p$, 
leaving numerous terms in the summations $\sum_{p}$ and $\sum_{q}$ of Eq.~\eqref{eqn:general_hessian}, 
thereby diminishing the flattening effect and complicating the loss landscape.
In our architecture, while the positional bias terms ($\bm{b}'_{h,w}$) are shared across multiple QNNs through overlapping patches, they affect only a limited, local number of patches. 
Therefore, the summation does not scale with $n_\text{qc}$, and the elements of the Hessian matrix corresponding to these bias terms remain effectively suppressed by the factors of $1/n_\text{qc}$ and $1/n_\text{qc}^2$.
Moreover, the total number of these positional bias terms ($14 \times 14 = 196$) is negligibly small compared to the vast number of independent variational parameters in the QNNs (ranging from approximately $10{,}000$ to $155{,}000$). 
Because of this tiny proportion, the sharing of these bias terms does not disrupt the global flattening effect on the loss landscape.
}

\added{
On the other hand, for the independent QNN parameters in our distributed architecture, 
assuming that $\theta_k$ belongs strictly to the $p_k$-th QNN ($\theta_k \in \bm{\Phi}_{p_k}$) and $\theta_j$ to the $p_j$-th QNN ($\theta_j \in \bm{\Phi}_{p_j}$), 
the derivatives $\frac{\partial \bm{y}_{i,p}}{\partial \theta_k}$ are zero for all $p \neq p_k$,
because the output $\bm{y}_{i,p}$ depends solely on its own parameter set $\bm{\Phi}_p$. 
Consequently, Eq.~\eqref{eqn:general_hessian} can be simply described by:
\begin{equation}
    \frac{\partial^2 C}{\partial \theta_j \partial \theta_k} = 
    \left( \frac{1}{n_\text{qc}} \right)^2 
    \left( \frac{\partial \bm{y}_{i,p_j}^\top}{\partial \theta_j} \right) 
    \left( \frac{\partial^2 C}{\partial \bar{\bm{y}}_i \partial \bar{\bm{y}}_i^\top} \right) 
    \left( \frac{\partial \bm{y}_{i,p_k}}{\partial \theta_k} \right) 
    + \frac{1}{n_\text{qc}} \left( \frac{\partial C}{\partial \bar{\bm{y}}_i^\top} \right) 
    \left( \frac{\partial^2 \bm{y}_{i,p_k}}{\partial \theta_j \partial \theta_k} \right).
    \label{eqn:simplified_hessian}
\end{equation}
Note that the mixed second derivative in the second term, $\frac{\partial^2 \bm{y}_{i,p_k}}{\partial \theta_j \partial \theta_k}$, 
survives if and only if both parameters belong to the same independent QNN ($p_j = p_k$). 
When $\theta_j$ and $\theta_k$ belong to different QNNs, 
this cross-derivative term is identically zero, 
making the second term of the Hessian matrix effectively block-diagonal with respect to the QNN parameters.
This would allow us to facilitate optimizing parameters during training. 
From the theoretical analysis,
we can see that our distributed model acts as an implicit structural regularization, 
effectively flattening the loss landscape and mitigating the trainability issues associated with deep QNNs. 
}
%
%

\subsection{Visualizing Loss Landscapes}\label{methods:vis}
To visualize loss landscapes and corresponding optimization trajectories, 
we employ a method adapted from Ref.~\cite{li2018visualizing}. 
In this study, we determine the visualization axes by applying principal component analysis (PCA) to the trajectory of the model's parameters recorded during optimization. 
Specifically, let $\bm{q}^{(i)}=({\bm{\Phi}_0^{(i)}}^\top, \ldots, {\bm{\Phi}_{n_\text{qc}-1}^{(i)}}^\top )^\top$ represent the concatenated vector of all variational parameters from all $n_\text{qc}$ QNNs at the end of the $i$-th optimization epoch. 
We construct a matrix $\bm{Q}=[{\bm{q}^{(\text{init})}}^\top; {\bm{q}^{(0)}}^\top; {\bm{q}^{(1)}}^\top; \ldots; {\bm{q}^{(n_\text{epoch}-1)}}^\top]$, 
where $\bm{q}^{(\text{init})}$ represents the parameter vector at initialization 
and $\bm{q}^{(i)}$ at the end of epochs $i=0, 1, \ldots, n_\text{epoch}-1$. 
We apply PCA to $\bm{Q}$ and select the first and second principal components, which define the orthogonal basis for the 2D visualization plane. 
To construct contour plots of the loss landscape, we first determined the range of the optimization trajectory projected onto the 2D visualization plane. 
We then compute the loss function values on a $20 \times 20$ grid of points spanning a region that covers the trajectory range with a suitable margin along each axis. 
To visualize the optimization trajectory, 
we project the sequence of recorded parameter vectors (i.e., $\bm{q}^{(\text{init})}$ and $\bm{q}^{(i)}$ for $i=0, \ldots, n_\text{epoch}-1$) onto this 2D visualization plane. 
This visualization technique enables a qualitative investigation into how the number of parameters and patches affects the geometric features of the loss landscape.

\section{Numerical Experiments}\label{sec:numerical_experiments}
In this section, we present numerical experiments to investigate how the number of variational parameters and the number of patches affect classification performance and loss landscape geometry of our distributed QNN model. 

\subsection{Setup}\label{subsec:settings}
Our numerical experiments are conducted using the MNIST dataset, consisting of $60{,}000$ training and $10{,}000$ test images.
Each image has $28 \times 28$ pixels with grayscale values ranging from $0$ to $255$, representing handwritten digits from $0$ to $9$. 
Directly using the original $28 \times 28$ images for loss landscape visualization and Hessian analysis is computationally expensive due to the high memory and processing demands. 

To reduce the computational costs, 
we downsample the images to $M\times M= 14 \times 14$ pixels by applying a $2 \times 2$ average pooling operation. 
Subsequently, the pixel values of these downsampled images are normalized into the range $[0,\pi/4]$ to serve as input angles for the QNNs. 
These preprocessed values correspond to the $x_{i,h,w}^\prime$ described in Section~\ref{methods:distqnns}. 
Additionally, we incorporate the learnable positional bias terms $b_{h,w}^\prime$, initialized to zero. 

Next, we partition each $14 \times 14$ preprocessed image into overlapping $P\times P=8 \times 8$ patches using strides $D\in\{6,3,2\}$. 
This process yields $n_\text{qc}\in\{4,9,16\}$ patches per image, respectively, 
corresponding to $L\in\{2,3,4\}$ patches along each dimension. 
Each patch is then assigned to one of the $n_\text{qc}$ individual QNNs. 

Each of the $n_\text{qc}$ QNN employs an $n=8$ qubit hardware-efficient Ansatz, shown in Figure~\ref{fig:qc}, 
and is detailed in Appendix~\ref{appendix:ansatz}. 
We initialize the variational parameters (rotation angles) $\bm{\Phi}_p$ for all QNNs by drawing values uniformly from the interval $[0,\pi]$. 
Both the parameters $\bm{\Phi}_p$ (for all $p$) and the bias terms $\bm{b}'_{h,w}$ are treated as trainable parameters and optimized simultaneously using the Adam optimizer~\cite{kingma2014adam} with a cosine annealing scheduler. 
The initial learning rate for the Adam optimizer was set according to the Ansatz depth $d$: 
$1.0\times10^{-2}$ (for $d=50$), $5.0\times10^{-3}$ (for $d=100$), 
$2.5\times10^{-3}$ (for $d=150$), and $1.0\times10^{-3}$ (for $d=200$). 
For classification, each QNN computes expectation values for a set of $n_\text{class}=10$ observables $\{O_k\}_{k=0}^9=\{X_0,\ldots,X_4,Z_0,\ldots,Z_4\}$. 
These expectation values $\bm{y}_{i,p}$ are averaged to produce $\overline{\bm{y}}_i$, as in Eq.~\eqref{eqn:averaged_outputs}. 
This averaged vector is scaled by a fixed constant value $c=100$ 
and then is passed through the Softmax function to yield class probabilities, 
as in Eq.~\eqref{eqn:class_prob}. 
The model is trained to minimize the cross-entropy loss. 
The training proceeds for $50$ epochs with a mini-batch size of $1{,}000$. 
Our classical simulations were performed using TorchQuantum~\cite{torchquantum2024}. 
After the optimization, 
we visualize the loss landscapes using the method described in Section~\ref{methods:vis}, 
and compute the largest Hessian eigenvalues at the minimum training loss.

\subsection{Results}\label{subsec:results}
\subsubsection{\added{Classification Performance and Generalization}}
As we described in Section~\ref{subsec:settings}, 
our numerical experiments employ a cosine annealing scheduler. 
This scheduling strategy means that 
the parameters at the final epoch may not coincide with those yielding 
the minimum training loss, as illustrated in Fig.~\ref{fig:train_test_loss}. 
Consequently, we report the test loss and test accuracy at the epoch that achieved the minimum training loss, presented in Table~\ref{table:loss_acc}. 
Table~\ref{table:nparams} summarizes the number of variational parameters for each configuration. 
A general trend observed from these tables is that increasing the number of parameters (via depth $d$, described in Appendix~\ref{appendix:ansatz}) or the number of patches ($n_\text{qc}$) tends to improve generalization performance. 

First, we analyze the impact of increasing the number of patches $n_\text{qc}$
while keeping the total number of parameters approximately constant. 
For instance, comparing configurations with roughly $20{,}000$ parameters, 
the configuration with $n_\text{qc}=9$ patches $(d=50)$ achieved a lower test loss than the one with $n_\text{qc}=4$ patches $(d=100)$. 
Similarly, comparing configurations with roughly $40{,}000$ parameters ($n_\text{qc}=4$ patches, $d=200$; $n_\text{qc}=9$ patches, $d=100$; $n_\text{qc}=16$ patches, $d=50$), 
the test loss \added{generally} decreased as the number of patches increased: 
$n_\text{qc}=16$ patches ($d=50$) \added{$\approx$} \deleted{$<$} $n_\text{qc}=9$ patches ($d=100$) $<$ $n_\text{qc}=4$ patches ($d=200$). 
\added{
Specifically, the performance improves when transitioning from $n_\text{qc}=4$ to $n_\text{qc}=9$ or $16$. 
While the performance between $n_\text{qc}=9$ ($d=100$) and $n_\text{qc}=16$ ($d=50$) is comparable, both significantly outperform the $n_\text{qc}=4$ ($d=200$) configuration. 
}
These results indicate that increasing the number of patches enhances generalization performance for a comparable number of parameters. 
\added{
Crucially, this performance enhancement is particularly substantial when transitioning away from the small patch configuration, 
where the first Hessian term dominates and complicates the optimization landscape. 
}

Next, we examine the effect of increasing the number of parameters for a fixed number of patches. 
For the $n_\text{qc}=4$ patch configuration, the test loss decreases as $d$ increases from $50$ to $150$, but slightly increases at $d=200$. 
This non-monotonic behavior suggests that while increasing the number of parameters can initially improve performance, 
\added{the model} \deleted{there} might encounter overfitting or optimization challenges, possibly related to a sharp loss landscape around the minimum. 
For the $n_\text{qc}=9$ patch configuration, the test loss consistently decreases as 
$d$ increases from $50$ to $200$. 
However, for the $n_\text{qc}=16$ patch configuration, the test loss remains almost constant despite the increase in the number of parameters, 
implying that the model's capacity exceeds what can be learned from the given data. 
These observations indicate that while increasing \added{the number of} parameters is generally beneficial, their effectiveness diminishes, potentially due to overfitting, optimization challenges, or reaching a data-limited regime hindering the effective use of additional parameters.

\subsubsection{\added{Loss Landscape Geometry and Hessian Analysis}}
\begin{figure}
     \centering
     \begin{subfigure}[b]{0.23\textwidth}
         \centering
         \includegraphics[width=\textwidth]{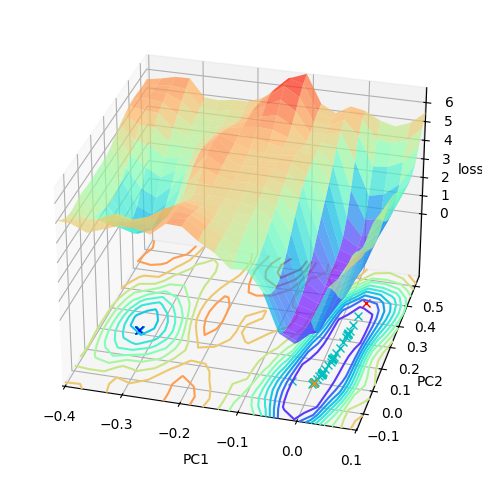}
         \caption{$n_\text{qc}=4$, $d=50$}
         \label{fig:4qnns50}
     \end{subfigure}
     \hfill
     \begin{subfigure}[b]{0.23\textwidth}
         \centering
         \includegraphics[width=\textwidth]{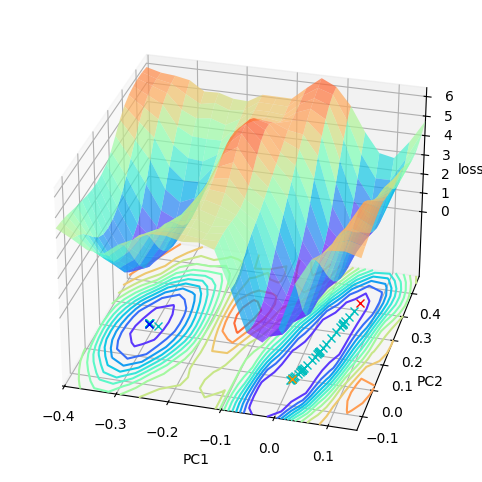}
         \caption{$n_\text{qc}=4$, $d=100$}
         \label{fig:4qnns100}
     \end{subfigure}
     \hfill
     \begin{subfigure}[b]{0.23\textwidth}
         \centering
         \includegraphics[width=\textwidth]{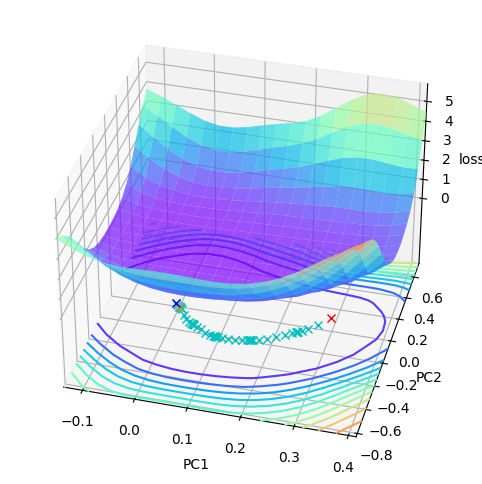}
         \caption{$n_\text{qc}=4$, $d=150$}
         \label{fig:4qnns150}
     \end{subfigure}
     \hfill
     \begin{subfigure}[b]{0.23\textwidth}
         \centering
         \includegraphics[width=\textwidth]{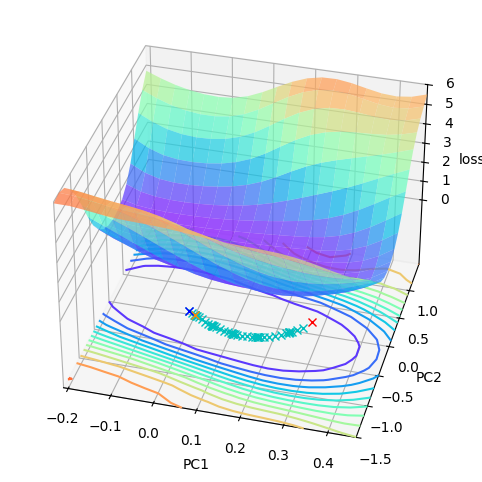}
         \caption{$n_\text{qc}=4$, $d=200$}
         \label{fig:4qnns200}
     \end{subfigure}
     \\
     \begin{subfigure}[b]{0.23\textwidth}
         \centering
         \includegraphics[width=\textwidth]{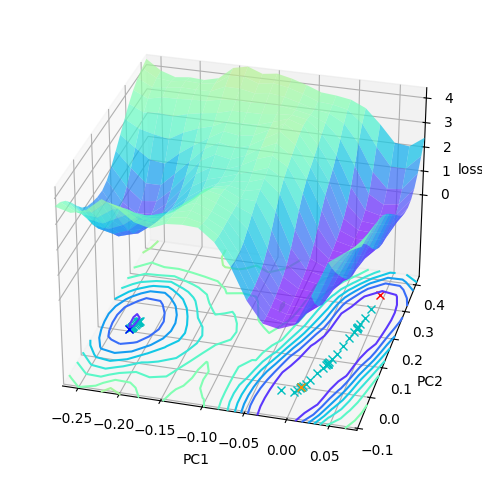}
         \caption{$n_\text{qc}=9$, $d=50$}
         \label{fig:9qnns50}
     \end{subfigure}
     \hfill
     \begin{subfigure}[b]{0.23\textwidth}
         \centering
         \includegraphics[width=\textwidth]{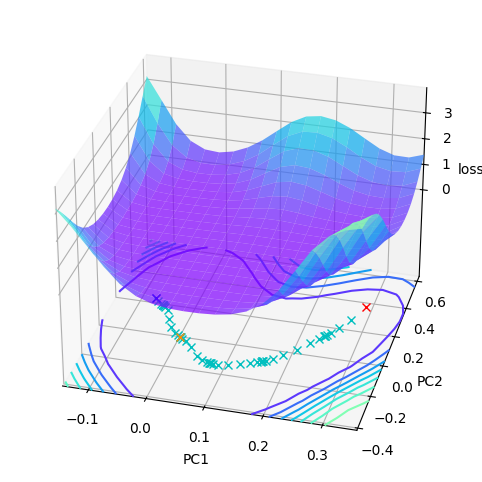}
         \caption{$n_\text{qc}=9$, $d=100$}
         \label{fig:9qnns100}
     \end{subfigure}
     \hfill
     \begin{subfigure}[b]{0.23\textwidth}
         \centering
         \includegraphics[width=\textwidth]{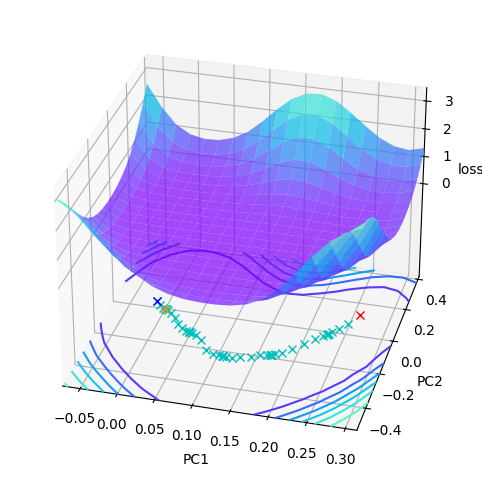}
         \caption{$n_\text{qc}=9$, $d=150$}
         \label{fig:9qnns150}
     \end{subfigure}
     \hfill
     \begin{subfigure}[b]{0.23\textwidth}
         \centering
         \includegraphics[width=\textwidth]{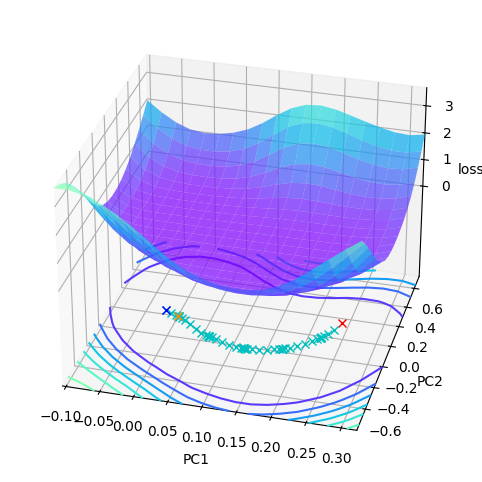}
         \caption{$n_\text{qc}=9$, $d=200$}
         \label{fig:9qnns200}
     \end{subfigure}    
     \\
     \begin{subfigure}[b]{0.23\textwidth}
         \centering
         \includegraphics[width=\textwidth]{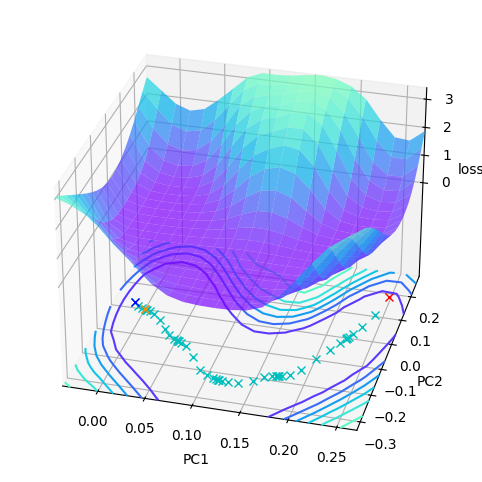}
         \caption{$n_\text{qc}=16$, $d=50$}
         \label{fig:16qnns50}
     \end{subfigure}
     \hfill
     \begin{subfigure}[b]{0.23\textwidth}
         \centering
         \includegraphics[width=\textwidth]{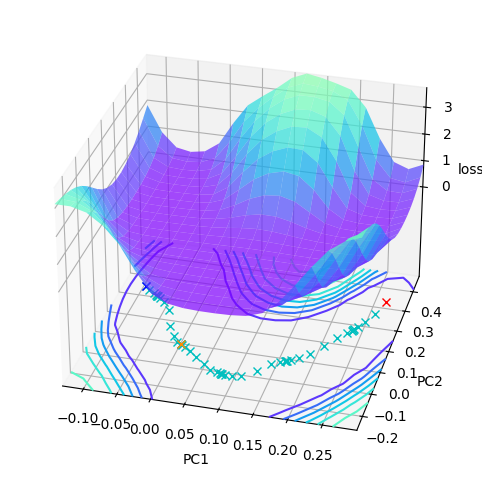}
         \caption{$n_\text{qc}=16$, $d=100$}
         \label{fig:16qnns100}
     \end{subfigure}
     \hfill
     \begin{subfigure}[b]{0.23\textwidth}
         \centering
         \includegraphics[width=\textwidth]{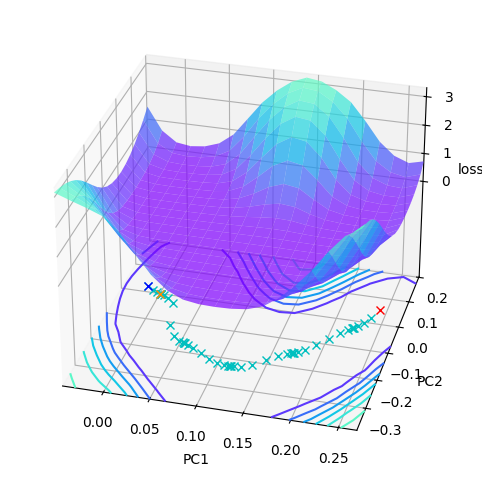}
         \caption{$n_\text{qc}=16$, $d=150$}
         \label{fig:16qnns150}
     \end{subfigure}
     \hfill
     \begin{subfigure}[b]{0.23\textwidth}
         \centering
         \includegraphics[width=\textwidth]{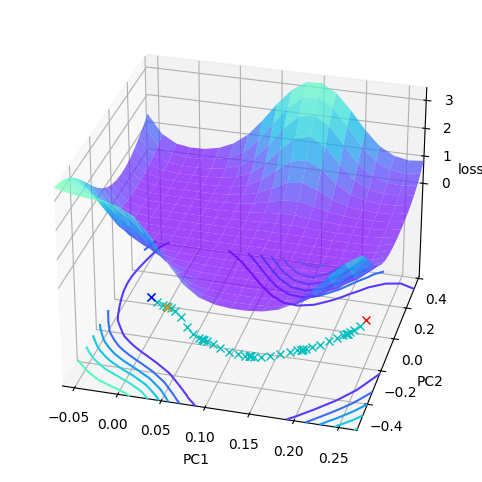}
         \caption{$n_\text{qc}=16$, $d=200$}
         \label{fig:16qnns200}
     \end{subfigure}    
    \caption{Visualization of training loss landscapes and optimization trajectories for each model:\\
    This figure illustrates the training loss landscapes and the optimization trajectories, 
    projected onto the 2D plane defined by the first two principal components of the model parameters (the first and the second principal components are denoted by PC1 and PC2, respectively). 
    The optimization trajectory starts at the position marked by a red cross and ends at the dark blue cross. 
    The orange cross marks the position of the minimum training loss during optimization. 
    The contour lines depict levels of the loss function at intervals of $0.5$.
    }
    \label{fig:losslandscapes_overview}
\end{figure}

To further understand these performance differences, we analyze the loss landscape geometry by visualizing it and computing the largest Hessian eigenvalue $\lambda_\text{max}$ using PyHessian~\cite{yao2020pyhessian}. 
A smaller \added{value for the} largest Hessian eigenvalue $\lambda_\text{max}$ indicates a flatter minimum, a characteristic often associated with improved generalization performance~\cite{keskar2016large}.
Visualizations in Figure~\ref{fig:losslandscapes_overview} suggest that 
for relatively small parameter counts, optimization trajectories may explore multiple loss basins (e.g., for $n_\text{qc}=4$ patches with $d=50$ and $d=100$, and $n_\text{qc}=9$ patches with $d=50$). 
In contrast, as the number of parameters increases, the trajectories tend to remain within a single basin. 
The visualizations also qualitatively suggest that increasing the number of parameters leads to sharper loss landscapes. 
For example, the visualized loss surface near the minimum found for the $(n_\text{qc}=4, d=200)$ configuration appears sharper than that for the $(n_\text{qc}=4, d=150)$ configuration. 
This visual observation is quantitatively confirmed by the largest Hessian eigenvalues $\lambda_\text{max}$ presented in Table~\ref{table:top1_hessian}. 
For $n_\text{qc}=4$ patch configuration, the largest eigenvalue increases from $25.39$ $(d=150)$ to $55.23$ $(d=200)$, indicating that increasing the \added{number of} parameters for this fixed patch number leads to sharper minima. 
Similarly, for $n_\text{qc}=9$ and $n_\text{qc}=16$ patch configurations, 
the largest eigenvalue increases as the number of parameters increases. 
Importantly, it is observed that the increase in the largest Hessian eigenvalue $\lambda_\text{max}$ accompanying the growth of parameter counts is mitigated as the number of patches increases. 

Building upon this observation, we now focus on assessing the impact of the number of patches by comparing configurations with similar parameter counts.
This analysis crucially confirms that increasing the number of patches mitigates the landscape sharpening effect typically associated with increased model depth. 
For models with approximately $20{,}000$ parameters, 
the largest Hessian eigenvalue $\lambda_\text{max}$ at the minimum decreased substantially from $22.54$ ($n_\text{qc}=4$, $d=100$) to $4.387$ ($n_\text{qc}=9$, $d=50$). 
A similar trend was observed for models with approximately $40{,}000$ parameters:
the largest Hessian eigenvalue $\lambda_\text{max}$ decreased dramatically from $55.23$ ($n_\text{qc}=4$, $d=200$) to $5.953$ ($n_\text{qc}=9$, $d=100$) and further to $3.340$ ($n_\text{qc}=16$,  $d=50$) as the number of patches increased. 
These results indicate that for a comparable number of parameters, 
employing more patches leads to a significantly flatter loss landscape around minima.
This flattening effect suggests a form of implicit regularization, which is often associated with improved generalization performance. 
Furthermore, even the larger model tested ($n_\text{qc}=16$, $d=200$\added{, with} approximately $155{,}000$ parameters) exhibited a largest Hessian eigenvalue $\lambda_\text{max}$ of only $6.688$. 
This value is notably smaller than the $\lambda_\text{max}$ observed for the sharpest model ($\lambda_\text{max}=55.23$ for $n_\text{qc}=4$, $d=200$), despite the large increase in the number of parameters, further supporting the significant flattening effect of using a high number of patches. 

\added{
These empirical observations are highly consistent with the mathematical implications derived in Section~\ref{methods:theory}. 
When $n_\text{qc}$ is small and training loss is almost zero, Eq.~\eqref{eqn:simplified_hessian} implies that the first Hessian term dominates the Hessian matrix elements. 
Actually, comparing the largest Hessian eigenvalue for $n_\text{qc}=4, 9$ in Table~\ref{table:top1_hessian}, 
the numerical results show that the largest Hessian eigenvalue at $n_\text{qc}=9$ decreased to approximately $1/5$ to $1/4$ of that at $n_\text{qc}=4$. 
This reduction almost matches the theoretical scaling ratio of the quadratic first term  in Eq.~\eqref{eqn:simplified_hessian}, 
which shrinks by a factor of $(1/9^2) / (1/4^2) = 16/81 \approx 1/5$. 
This quantitative agreement strongly implies that the first term ($1/n_\text{qc}^2$) dominates the Hessian matrix elements in the small $n_\text{qc}$ region. 
When $n_\text{qc}$ is large, the first term (scaling as $1/n_\text{qc}^2$) decays rapidly, 
and the Hessian would typically be dominated by the second term (scaling as $1/n_\text{qc}$) during training. 
However, when the training loss is close to zero, the second term becomes almost zero. 
This is because, for the cross-entropy loss with the Softmax function, 
the gradient of the loss with respect to the aggregated output, $\partial C / \partial \bar{\boldsymbol{y}}_i^\top$, is proportional to the prediction error, which approaches zero as the loss vanishes. 
This structural property theoretically supports the globally flattened loss landscape observed in our numerical experiments. 
}

In summary, our results highlight the distinct and complementary roles of parameters and patches in this distributed QNN architecture. 
Increasing the number of parameters allows the model to potentially reach lower loss values, but can also lead to sharper minima. 
Conversely, increasing the number of patches promotes flatter minima for a given number of parameters. 
This flattening effect, acting as a form of implicit regularization inherent to this distributed patch-based approach, likely facilitates optimization and is strongly correlated with the improvements in generalization performance.

\begin{table}[ht]
\caption{The test loss and test accuracy 
at the epoch of minimum training loss}\label{table:loss_acc}
\begin{tabular*}{\textwidth}{@{\extracolsep\fill}lcc cc cc cc}
\toprule%
& \multicolumn{2}{@{}c@{}}{d=50} & \multicolumn{2}{@{}c@{}}{d=100} 
& \multicolumn{2}{@{}c@{}}{d=150} & \multicolumn{2}{@{}c@{}}{d=200} \\ 
\cmidrule{2-3}\cmidrule{4-5}\cmidrule{6-7}\cmidrule{8-9}%
          & Loss & Acc & Loss & Acc & Loss & Acc & Loss & Acc \\
\midrule
$n_\text{qc}=4$ & 0.06607 & 0.9798  & 0.06384 & 0.9818 & $\bm{0.05608}$ & $\bm{0.9822}$ & 0.06061 & 0.9815 \\
$n_\text{qc}=9$ & 0.05520 & 0.9823  & 0.05369 & 0.9832 & 0.05179 & $\bm{0.9844}$ & $\bm{0.04813}$ & 0.9839  \\
$n_\text{qc}=16$ & 0.04957 & 0.9840 & $\bm{0.04916}$ & 0.9837 & 0.04973 & 0.9848 & 0.05003 & $\bm{0.9854}$ \\
\botrule
\end{tabular*}
\footnotetext{
Table~\ref{table:loss_acc} shows the test loss and test accuracy at the minimum training loss 
using our model with $n_\text{qc}=4$, $9$, or $16$, and $d=50$, $100$, $150$, or $200$ in the entangling layers.
Note that we used the MNIST dataset, which was processed by $2 \times 2$ average pooling, resulting in a $14 \times 14$-sized dataset.
}
\end{table}

\begin{table}[ht]
\caption{The number of variational parameters in QNNs}\label{table:nparams}%
\begin{tabular}{l || llll}
\toprule
          & $d=50$ & $d=100$ & $d=150$ & $d=200$ \\
\midrule
$n_\text{qc}=4$ &  $10{,}112$  & $19{,}712$  & $29{,}312$  & $38{,}912$ \\
$n_\text{qc}=9$ &  $22{,}752$  & $44{,}352$  & $65{,}952$  & $87{,}552$ \\
$n_\text{qc}=16$ &  $40{,}448$  & $78{,}848$  & $117{,}248$ & $155{,}648$ \\
\botrule
\end{tabular}
\footnotetext{
Table~\ref{table:nparams} displays the number of variational parameters in our distributed QNN model.
}
\end{table}

\begin{table}[ht]
\caption{The largest Hessian eigenvalue}\label{table:top1_hessian}%
\begin{tabular}{l || llll}
\toprule
          & $d=50$  & $d=100$ & $d=150$ & $d=200$ \\
\midrule
$n_\text{qc}=4$ &  $15.98$  & $22.54$  & $25.39$ & $55.23$ \\
$n_\text{qc}=9$ &  $4.482$  & $5.953$  & $6.091$ & $10.01$ \\
$n_\text{qc}=16$ &  $3.340$  & $4.172$  & $4.515$ & $6.688$ \\
\botrule
\end{tabular}
\footnotetext{
Table~\ref{table:top1_hessian} represents the largest Hessian eigenvalue evaluated at the minimum training loss. 
}
\end{table}

\subsubsection{\added{Statistical validation and analysis}}\label{subsec:statistical_validation}
\added{
In this subsection, we perform numerical experiments to rigorously validate the observations discussed in the previous subsection, using a recent fast variational quantum circuit simulator \cite{kawase2026fast}. 
Specifically, we computed the test loss, test accuracy, and the largest Hessian eigenvalue with their means and standard deviations, evaluated at the minimum training loss across five different random seeds. 
The results of the generalization performance are shown in Table~\ref{table:loss_acc_multi_seeds}. 
From the table, we observe that 
the test loss and accuracy consistently improve as the number of patches increases for a fixed circuit depth $d$. 
Note that the performance improves significantly when the number of patches transitions from $n_\text{qc}=4$ to $n_\text{qc}=9$, 
whereas the improvement is only slight when further increasing from $n_\text{qc}=9$ to $n_\text{qc}=16$. 
In addition, for each fixed number of patches, the performance improves as the depth increases up to $d=150$, 
but it slightly worsens at $d=200$.
This phenomenon implies that our models are transitioning out of the under-parameterized regime and entering what is classically recognized as the overfitting regime, which likely lies between the under-parameterized and the over-parameterized regimes. 
}

\added{
Furthermore, Table~\ref{table:top1_hessian_multi_seeds} reports the statistical results of the largest Hessian eigenvalues evaluated at the minimum training losses across multiple seeds. 
Similar to the single-seed results in Table~\ref{table:top1_hessian}, 
Table~\ref{table:top1_hessian_multi_seeds} shows that for a fixed number of patches $n_\text{qc}$, 
the largest Hessian eigenvalue grows as the depth of the quantum circuit ($d$) increases. 
Conversely, for a fixed depth $d$, increasing the number of patches consistently suppresses the largest Hessian eigenvalue. 
Notably, the reduction of the largest Hessian eigenvalue is significantly larger 
when transitioning from $n_\text{qc}=4$ to $n_\text{qc}=9$ than from $n_\text{qc}=9$ to $n_\text{qc}=16$. 
Specifically, when $n_\text{qc}$ transitions from $4$ to $9$, the largest Hessian eigenvalues decrease to approximately $1/4$ to $1/5$ of their original values, 
confirming that the first term in Eq.~\eqref{eqn:simplified_hessian} dominates the sharpness of the loss landscape, especially when $n_\text{qc}$ is small. 
}

\begin{figure}
     \centering
     \includegraphics[width=0.95\textwidth]{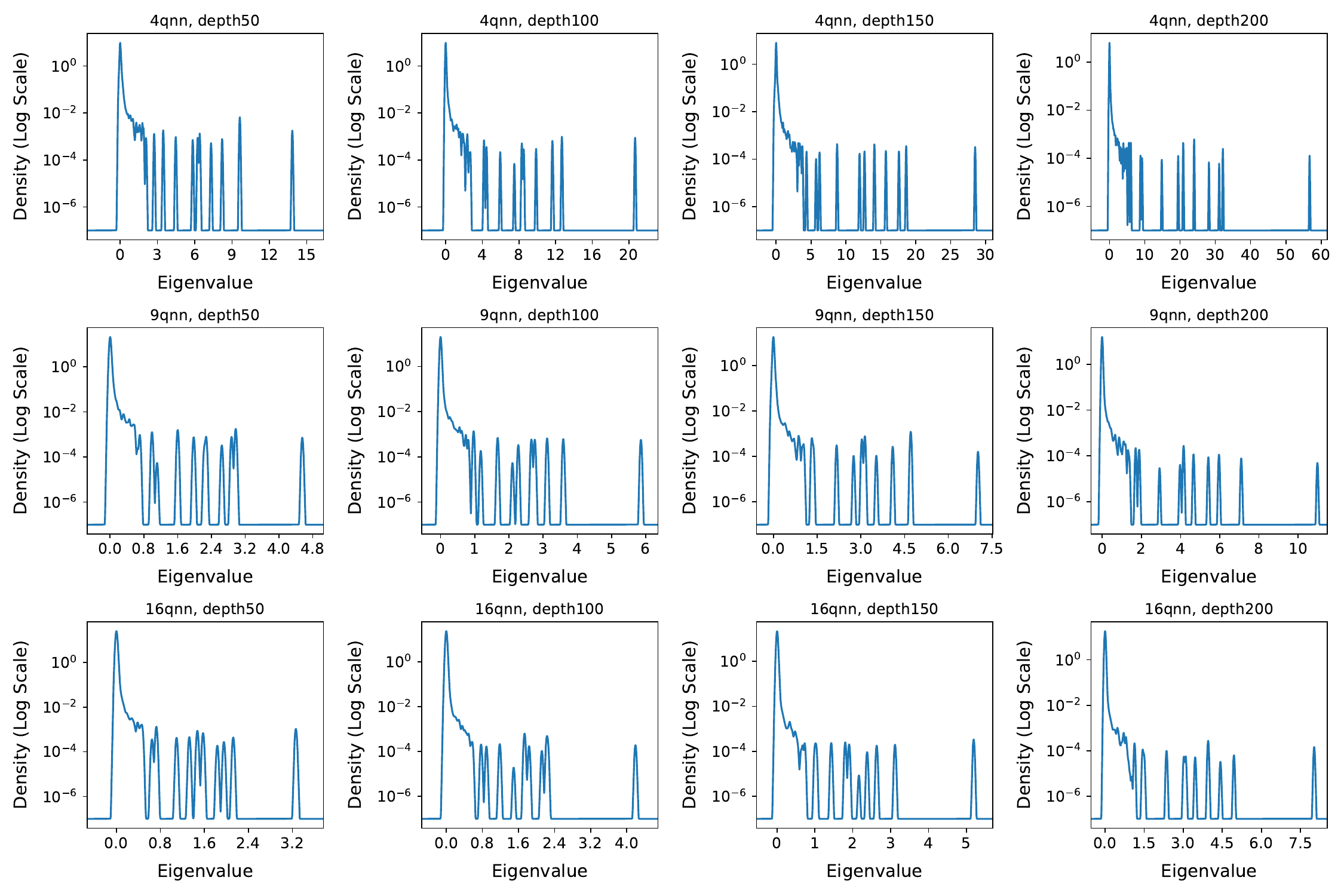}
     \caption{
    \added{
    Visualization of the Hessian eigenspectrum evaluated at the minimum training loss. \\
    The subplots display the distributions of Hessian eigenvalues for a representative random seed 
    across different numbers of patches ($n_{qc} \in \{4, 9, 16\}$, from top to bottom rows) 
    and circuit depths ($d \in \{50, 100, 150, 200\}$, from left to right columns). 
    The x-axis represents the eigenvalue, and the y-axis indicates the density on a logarithmic scale. 
    The spectra typically consist of a large bulk near zero and a small number of outlier spikes, approximately corresponding to the number of classes. 
    As the number of patches $n_{qc}$ increases, the entire eigenspectrum, including the bulk and the largest eigenvalue, is significantly compressed toward zero. 
    This visualizes and confirms the global flattening effect of our distributed patch architecture.
    }
     }
     \label{fig:density_plot}
\end{figure}

\added{
Fig.~\ref{fig:density_plot} shows the Hessian eigenspectrum for a representative seed. 
We can see that most of the eigenvalues have small values near zero, forming a bulk, 
while some eigenvalues separate from the bulk as outliers. 
Among these outliers, there are approximately $n_\text{class}$ distinct spikes. 
This phenomenon is often observed in classical deep learning for classification tasks \cite{sagun2016eigenvalues, papyan2018full}, 
particularly when highly expressive models achieve near-zero training loss while maintaining high generalization performance. 
This suggests that, as expected by Eq.~\eqref{eqn:simplified_hessian}, regardless of whether the model is classical or quantum, 
the Hessian eigenspectrum is dominated by the first Hessian term, which reflects the structure of the training data and the loss function. 
Furthermore, the largest Hessian eigenvalue deviates significantly from the rest of the eigenvalues, even among the outliers. 
As the circuit depth $d$ increases, the overall magnitudes of the Hessian eigenvalues increase, 
and the discrepancy between the largest eigenvalue and the others becomes significantly larger. 
Nevertheless, as the number of patches increases, 
not only the largest Hessian eigenvalue but also the entire eigenspectrum is suppressed toward zero. 
This suggests that increasing the number of patches effectively mitigates optimization difficulty by flattening the entire loss landscape geometry. 
}

\begin{table}[ht]
\centering
\caption{\added{
The test loss and test accuracy at the epoch of minimum training loss with multiple seeds
}}\label{table:loss_acc_multi_seeds}

\added{
\begin{tabular}{l cccc}
\toprule
& \multicolumn{2}{c}{d=50} & \multicolumn{2}{c}{d=100} \\ 
\cmidrule(lr){2-3}\cmidrule(lr){4-5}
          & Loss & Acc & Loss & Acc \\
\midrule
$n_\text{qc}=4$  & $0.06594 \pm 0.00507$ & $0.97923 \pm 0.00195$  & $0.06166 \pm 0.00152$ & $0.98127 \pm 0.00038(6)$ \\
$n_\text{qc}=9$  & $0.05502 \pm 0.00168$ & $0.98303 \pm 0.00061(1)$  & $0.05125 \pm 0.00208$ & $0.98444 \pm 0.00068(2)$ \\
$n_\text{qc}=16$ & $0.05362 \pm 0.00128$ & $0.98328 \pm 0.00039(6)$ & $0.04945 \pm 0.00213$ & $\bm{0.98478 \pm 0.00052(3)}$ \\
\botrule
\end{tabular}
}
\vspace{1.5em}

\added{
\begin{tabular}{l cccc}
\toprule
& \multicolumn{2}{c}{d=150} & \multicolumn{2}{c}{d=200} \\ 
\cmidrule(lr){2-3}\cmidrule(lr){4-5}
          & Loss & Acc & Loss & Acc \\
\midrule
$n_\text{qc}=4$  & $\bm{0.05690 \pm 0.00190}$ & $\bm{0.98195 \pm 0.00074(1)}$ & $0.06179 \pm 0.00365$ & $0.98103 \pm 0.00079(1)$ \\
$n_\text{qc}=9$  & $\bm{0.05094 \pm 0.00108}$ & $\bm{0.98464 \pm 0.00044(9)}$ & $0.05328 \pm 0.00265$ & $0.98333 \pm 0.00061(5)$ \\
$n_\text{qc}=16$ & $\bm{0.04798 \pm 0.00100}$ & $0.98452 \pm 0.00048(3)$ & $0.05025 \pm 0.00079(1)$ & $0.98434 \pm 0.00051(2)$ \\
\botrule
\end{tabular}
\footnotetext{
Similar to Table~\ref{table:loss_acc}, this table shows the test loss and accuracy with means and standard deviations for five different seeds at the minimum training loss. 
}
} 
\end{table}

\begin{table}[ht]
\caption{\added{
The largest Hessian eigenvalue for multiple seeds
}}\label{table:top1_hessian_multi_seeds}%
\added{
\begin{tabular}{l || llll}
\toprule
          & $d=50$  & $d=100$ & $d=150$ & $d=200$ \\
\midrule
$n_\text{qc}=4$ &  $19.00 \pm 4.012$  & $23.13 \pm 1.940$  & $29.74 \pm 2.244$ & $54.80 \pm 2.299$ \\
$n_\text{qc}=9$ &  $4.782 \pm 0.228$  & $6.241 \pm 0.466$  & $7.058 \pm 0.154$ & $11.16 \pm 0.684$ \\
$n_\text{qc}=16$ &  $3.195 \pm 0.100$  & $3.925 \pm 0.387$  & $5.343 \pm 0.204$ & $8.275 \pm 0.763$ \\
\botrule
\end{tabular}
\footnotetext{
This table represents the means and the standard deviations of largest Hessian eigenvalues evaluated at the minimum training loss for $5$ different seeds. 
}
} 
\end{table}

\section{Conclusion}\label{sec:conclusion}
In this study, we investigated how the number of parameters and the number of distributed patches affect the loss landscape geometry of a distributed QNN architecture for classical image classification. 
Specifically, we extracted the $8 \times 8$ overlapping local image patches from the MNIST dataset and used these patches as inputs to individual QNNs. 
The outputs from these QNNs were aggregated and passed through the Softmax function for classification. 
Through loss landscape visualization and \added{theoretical and empirical Hessian analyses} \deleted{Hessian analysis} at minima to characterize the loss landscape geometry, we identified two key findings. 

First\deleted{ly}, increasing the number of parameters enabled the models to \added{potentially} reach lower loss values but resulted in sharper minima, characterized by larger $\lambda_\text{max}$. 
While deeper minima are potentially beneficial, 
this increased sharpness can hinder optimization \added{stability} and negatively impact generalization. 

Second\deleted{ly}, and crucially, our \added{theoretical} analysis revealed that increasing the number of patches effectively mitigated this sharpening effect, 
acting as a form of implicit regularization and leading to significantly flatter minima \deleted{for a comparable number of parameters}. 
\added{We also empirically confirmed this landscape flattening effect. 
Especially when transitioning from a small to a moderately larger number of patches, 
the dominant Hessian term scales down by a factor of $1/n^2_\text{qc}$ as predicted by the theoretical analysis, 
causing a drastic reduction in landscape sharpness. }
This landscape flattening \added{effect} is strongly correlated with improved optimization stability and enhanced generalization performance, 
suggesting \added{that this distributed approach} \deleted{it} helps mitigate training difficulties often encountered in QNNs. 

\added{
In addition, by investigating the full Hessian eigenspectrum, 
we observed that our models exhibit a spectral structure similar to that of classical deep learning models. 
Specifically, the spectrum consists of a large bulk of eigenvalues concentrated near zero and a small number of outlier spikes, 
which roughly correspond to the number of classes. 
This finding provides valuable insight, suggesting that distributed QNNs share optimization characteristics with classical deep learning models. 
As future work, it is important to investigate whether the number of outlier spikes in the Hessian eigenspectrum of QNNs exactly corresponds to the number of classes, how this bulk and these spikes evolve during training, and how they affect generalization performance. 
}

\deleted{
These findings highlight that employing local feature patches distributed across multiple QNNs is a beneficial approach for quantum machine learning applied to classical data. 
This architectural choice appears effective in promoting flatter loss landscapes, 
leading to more stable training and better generalization. 
}

\added{
These findings offer practical implications for designing scalable QML models. 
For instance, our theoretical analysis implies that simply averaging the outputs (scaling by $1/n_\text{qc}$) decays the first Hessian term too rapidly ($1/n_\text{qc}^2$). 
To preserve an appropriate level of gradient magnitudes while increasing the number of patches, 
replacing the simple average with a scaling factor of $1/\sqrt{n_\text{qc}}$ for the aggregated output might be a highly effective architectural improvement. 
}

\added{
In addition, as future work, our theoretical Hessian equation motivates investigating the effect of cross-derivative terms 
by comparing the distributed QNN model with a single large QNN. 
The cross-derivative terms in the Hessian matrix between parameters of different QNNs are identically zero in the distributed QNN model
because the individual QNNs operate independently, whereas they would be generally non-zero in a single global QNN. 
Exploring how this structural difference impacts the optimization landscape presents a promising direction for future research. 
}

\deleted{
A primary limitation is our focus on the largest Hessian eigenvalue $\lambda_\text{max}$ due to computational costs. 
Future studies incorporating a full Hessian spectrum analysis could provide a more comprehensive understanding of the loss landscape characteristic of this distributed QNN architecture. 
Furthermore, investigating whether a similar local patch strategy could benefit other QML methods facing trainability issues, such as quantum kernel methods exhibiting concentration issues~\cite{thanasilp2024exponential}, presents an interesting avenue for future research. 
}

\added{
We acknowledge that our current results are based on noiseless classical simulations using the relatively simple MNIST dataset. 
Therefore, our study serves as an empirical and theoretical investigation that provides foundational insights into QNN trainability, 
rather than a direct demonstration of the practical utility of quantum machine learning models on current NISQ devices. 
Future studies incorporating noisy environments and applying the models to more complex datasets present promising avenues for future research.
}

\deleted{
In summary, this research contributes to the advancement of quantum machine learning 
by providing insights into the impacts of parameter count and data distribution via patches on loss landscape geometry. 
We demonstrate that a distributed local patch approach offers a practical strategy for promoting flatter minima, 
thereby mitigating optimization challenges and enhancing the performance of this distributed QNN architecture applied to classical data tasks. 
}

\added{
In summary, although developing techniques to effectively train a single large-scale global QNN remains an ideal goal for QML, 
our distributed local patch approach currently offers a powerful and scalable architectural choice. 
By providing insights into how parameter counts and data distribution via patches affect the loss landscape geometry, 
we demonstrate that this approach acts as an implicit regularization that promotes flatter minima. 
Consequently, it effectively mitigates optimization challenges and enhances the performance of QNNs applied to classical data tasks.
}

\backmatter

\section*{Declarations}

\bmhead{Acknowledgements}
This study was partially conducted using the Supermicro ARS-111GL-DNHR-LCC and NVIDIA Hopper H100 GPU (Miyabi-G) at Joint Center for Advanced High Performance Computing (JCAHPC). 

\bmhead{Funding}
This work was supported by JSPS KAKENHI Grant Number JP23K19954.

\bmhead{Conflict of interest/Competing interests}
The author declares no competing interests.

\bmhead{Data Availabitilty}
The data set used in this study is publicly available. 

\bmhead{Code Availabitilty}
The code used in this study is available at GitHub (\url{https://github.com/puyokw/qnns_with_patches}).

\bmhead{Author contribution}
Yoshiaki Kawase completed all tasks. 

\begin{appendices}

\section{Ansatz structure}\label{appendix:ansatz}
\begin{figure}
     \centering
     \begin{subfigure}[b]{\textwidth}
         \centering
         \includegraphics[width=\textwidth]{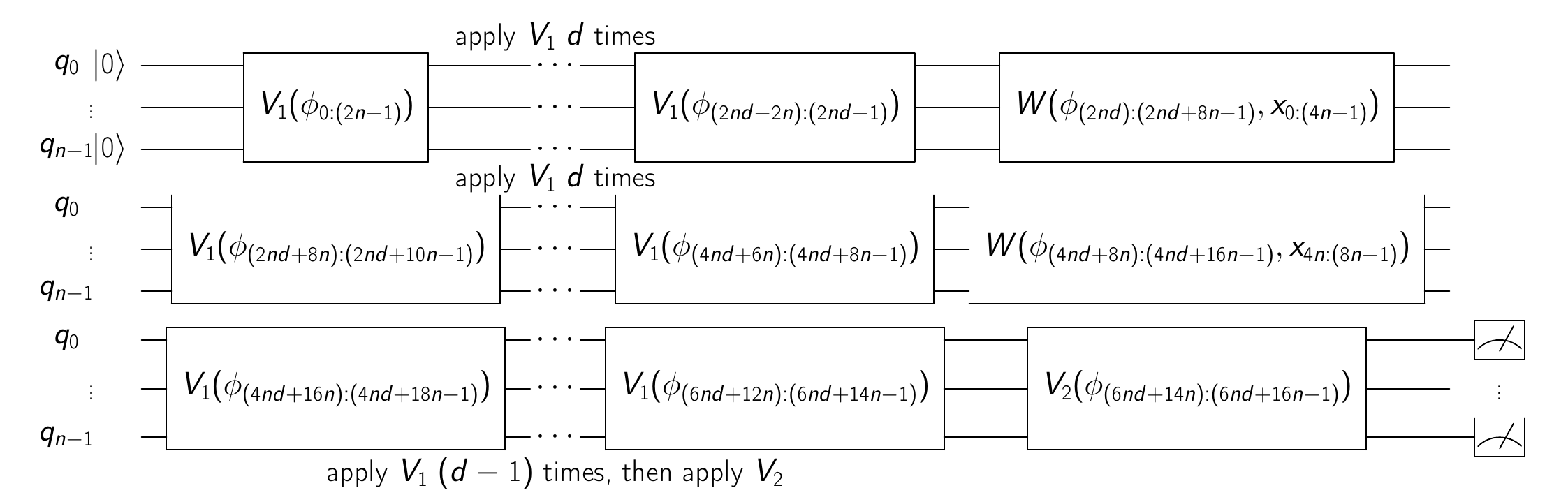}
         \caption{The overview of a quantum circuit we used in our numerical experiments}
         \label{fig:qc_overview}
     \end{subfigure}
     \hfill
     \begin{subfigure}[b]{\textwidth}
         \centering
         \includegraphics[width=\textwidth]{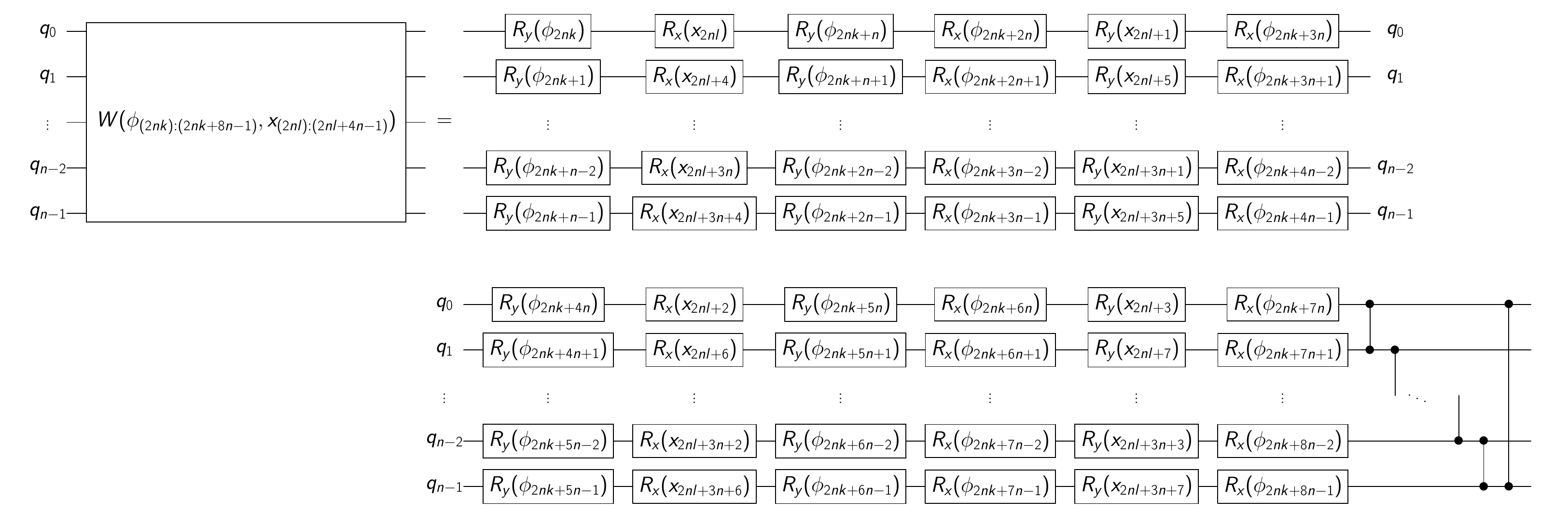}
         \caption{A data encoding block}
         \label{fig:qc_x}
     \end{subfigure}
     \hfill
     \begin{subfigure}[b]{0.54\textwidth}
         \centering
         \includegraphics[width=\textwidth]{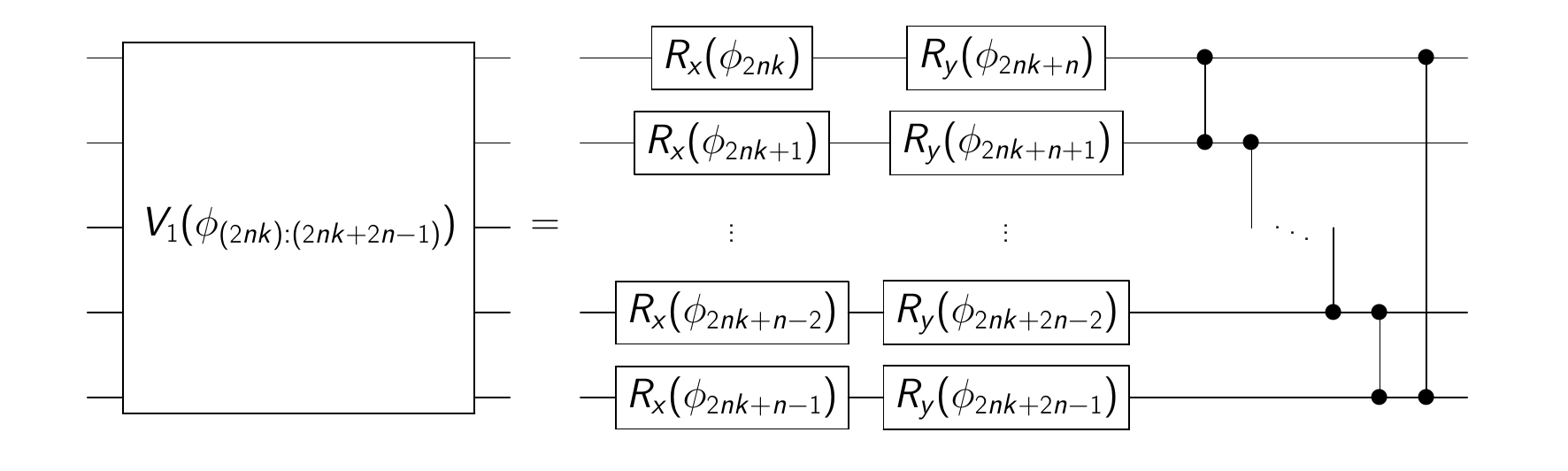}
         \caption{An entangling block}
         \label{fig:qc_phi}
    \end{subfigure}
     \hfill
     \begin{subfigure}[b]{0.44\textwidth}
         \centering
         \includegraphics[width=\textwidth]{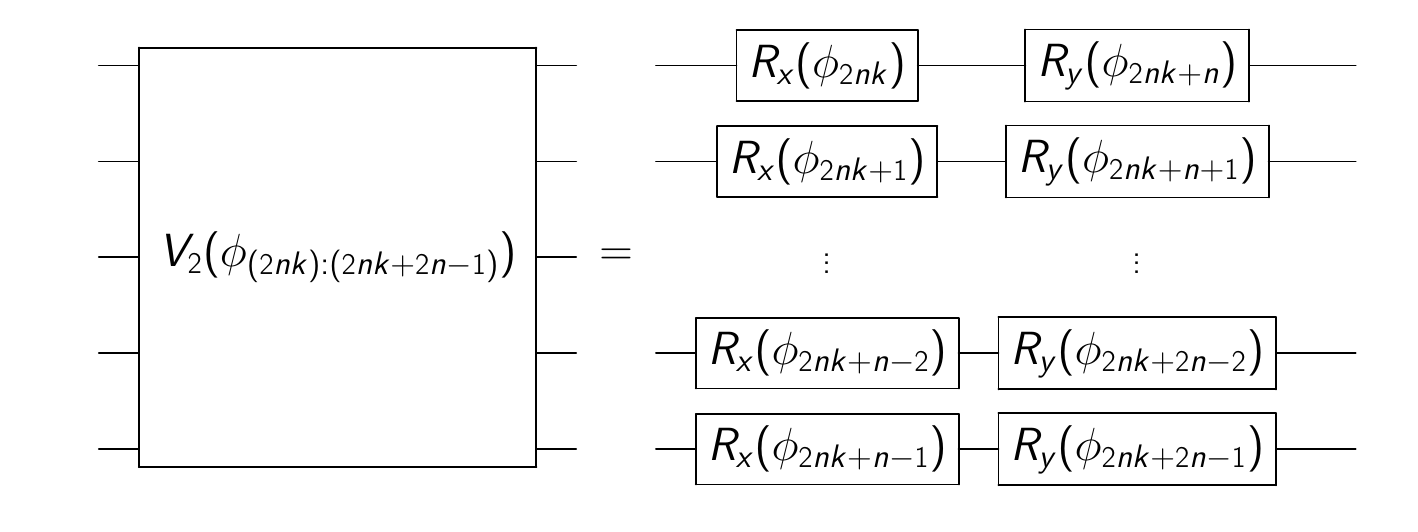}
         \caption{The final layer}
         \label{fig:qc_phi_final}
    \end{subfigure}
    \caption{
    Figure~\ref{fig:qc_overview} shows the overview of a quantum circuit we used in our numerical experiments. 
    Note that $d$ and $n$ represent the number of repetitions of the entangling block $V_1$ and the number of qubits. 
    In our numerical experiments, we set $n=8$, and $d=50, 100, 150$ or $200$.}
    \label{fig:qc}
\end{figure}

Here, we detail the Ansatz structure used in our numerical experiments. 
As we mentioned in Section~\ref{subsec:settings}, 
our Ansatz is an $n=8$ qubit hardware-efficient Ansatz, shown in Figure~\ref{fig:qc}, 
constructed from two types of blocks: a data encoding block and an entangling block. 

The data encoding block consists of two repetitions of a rotation sequence, followed by an entanglement layer. 
The rotation sequence includes six consecutive single qubit rotation layers ($R_y$, $R_x$, $R_y$, $R_x$, $R_y$, and $R_x$) applied to all $n=8$ qubits. 
Among these layers, the rotation angles for the first ($R_y$), third ($R_y$), fourth ($R_x$), and sixth ($R_x$) layers are trainable parameters, collectively part of $\bm{\Phi}_p$. 
The second ($R_x$) and fifth ($R_y$) rotation layers are used for angle encoding of input data. 
The entanglement layer applies $n$ CZ gates acting between adjacent qubits $i$ and $(i+1)$ $\pmod{n}$, assuming periodic boundary conditions. 

The entangling block consists of a layer of $R_x$ rotation gates and a layer of $R_y$ rotation gates, both with trainable parameters included in $\bm{\Phi}_p$, applied to all qubits, followed by a layer of $n$ CZ gates applied between adjacent qubits $i$ and $(i+1)$ $\pmod{n}$. 

The overall circuit architecture for the $p$-th QNN depends on a depth hyperparameter $d$ and\added{, 
similar to the data re-uploading architecture \cite{perez2020data} widely used to enhance the expressibility of quantum models \cite{jerbi2023quantum, schuld2021effect}, follows an interleaved sequence} 
\deleted{follows the sequence in this interleaved structure} of data encoding and trainable blocks: 
$d$ entangling blocks, one data encoding block, $d$ entangling blocks, one data encoding block, and $d$ entangling blocks. 
The circuit concludes with final $R_x$ and $R_y$ rotation layers, omitting the last CZ layer from the final entangling block.

The $64$ features of the input patch vector $\bm{x}_{i,p}$ are encoded using the second and fifth rotation layers within the two data encoding blocks present in the overall architecture. 
Recall that each data encoding block contains two repetitions of the rotation sequence, 
and within each sequence, the second ($R_x$)and fifth ($R_y$) layers are used for data encoding. 
Therefore, each block provides $2\times 2=4$ encoding layers, each acting on $n=8$ qubits. 
The two data encoding blocks in the overall architecture thus encode all $64$ features. 
The set of variational parameters $\bm{\Phi}_p$ includes all trainable rotation angles throughout the $p$-th QNN, excluding those used for data encoding.

\section{The contribution ratios of principal components}\label{appendix:cont_ratio}
This section discusses the contribution ratios of the principal components. 
Since our contour plots are based on the first two principal components, 
it is important to evaluate their cumulative ratio of those principal components. 
This evaluation confirms whether the 2D visualization adequately captures the optimization trajectories. 
As shown in Table~\ref{tab:cont_ratio}, 
the cumulative contribution ratio of the first two principal components exceeds 0.9000 for all models examined in this study. 
These high cumulative contribution ratios indicate that 
the vast majority of the variance within the optimization trajectories is represented by the first two principal components.
Therefore, our visualization approach provides a sufficiently accurate representation of the optimization process.

\begin{table}[ht]
\caption{The Contribution rate of Principal Components}\label{tab:cont_ratio}
\begin{tabular*}{\textwidth}{@{\extracolsep\fill}lcc cc cc cc}
\toprule%
& \multicolumn{2}{@{}c@{}}{$d=50$} & \multicolumn{2}{@{}c@{}}{$d=100$} 
& \multicolumn{2}{@{}c@{}}{$d=150$} & \multicolumn{2}{@{}c@{}}{$d=200$} \\ 
\cmidrule{2-3}\cmidrule{4-5}\cmidrule{6-7}\cmidrule{8-9}%
          & PC1 & PC2 & PC1 & PC2 & PC1 & PC2 & PC1 & PC2 \\
\midrule
$n_\text{qc}=4$ & 0.8753 & 0.0849  & 0.8295 & 0.1467 & 0.9386 & 0.0371 & 0.9578 & 0.0280 \\
$n_\text{qc}=9$ & 0.8880 & 0.0784  & 0.8489 & 0.0869 & 0.9217 & 0.0454 & 0.9586 & 0.0284  \\
$n_\text{qc}=16$ & 0.8585 & 0.0699 & 0.9024 & 0.0567 & 0.9410 & 0.0354 & 0.9690 & 0.0206 \\
\botrule
\end{tabular*}
\footnotetext{
Table~\ref{tab:cont_ratio} shows the contribution ratio of first and second principal components, denoted by PC1 and PC2 in this Table, 
using our model with $n_\text{qc}=4$, $9$, or $16$, and $d=50$, $100$, $150$, or $200$ in the entangling layers. 
}
\end{table}

\section{The training and test losses}\label{appendix:losses}
\setcounter{figure}{0}

\begin{figure}[ht]
     \centering
     \begin{subfigure}{0.24\textwidth}
         \centering
         \includegraphics[width=\textwidth]{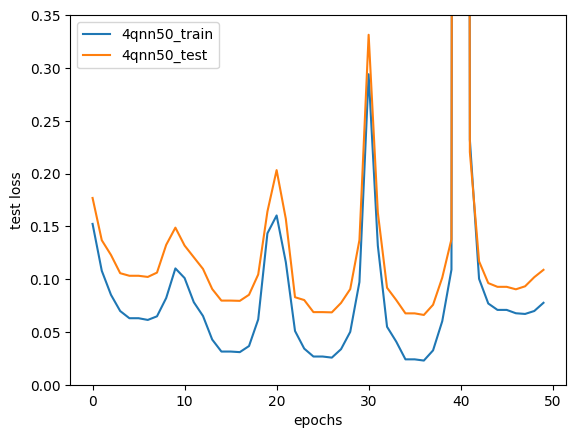}
         \caption{$n_\text{qc}=4$, $d=50$}
         \label{fig:4qnns50_loss}
     \end{subfigure}
     \hfill
     \begin{subfigure}{0.24\textwidth}
         \centering
         \includegraphics[width=\textwidth]{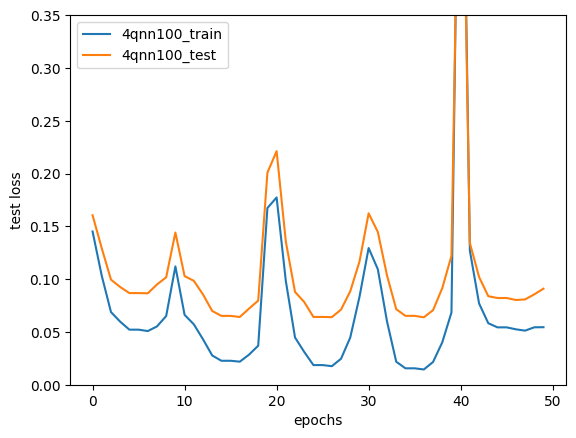}
         \caption{$n_\text{qc}=4$, $d=100$}
         \label{fig:4qnns100_loss}
     \end{subfigure}
     \hfill
     \begin{subfigure}{0.24\textwidth}
         \centering
         \includegraphics[width=\textwidth]{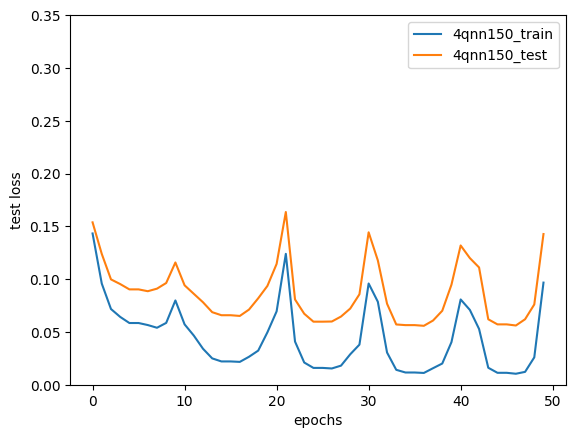}
         \caption{$n_\text{qc}=4$, $d=150$}
         \label{fig:4qnns150_loss}
     \end{subfigure}
     \hfill
     \begin{subfigure}{0.24\textwidth}
         \centering
         \includegraphics[width=\textwidth]{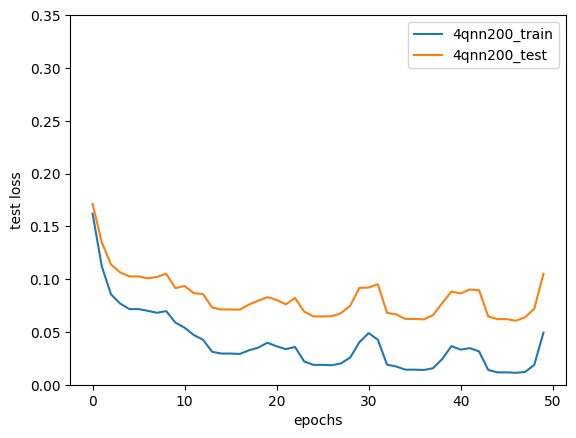}
         \caption{$n_\text{qc}=4$, $d=200$}
         \label{fig:4qnns200_loss}
     \end{subfigure}
     \\
     \begin{subfigure}{0.24\textwidth}
         \centering
         \includegraphics[width=\textwidth]{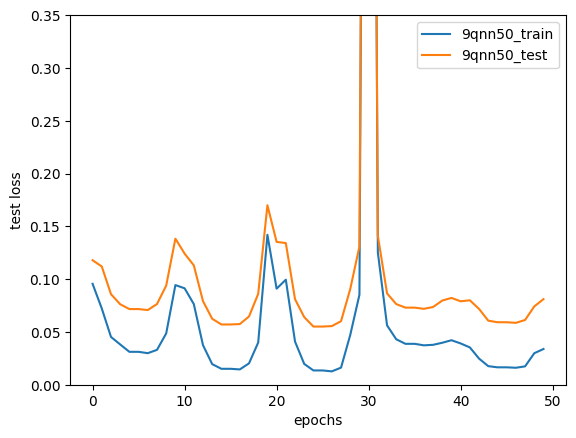}
         \caption{$n_\text{qc}=9$, $d=50$}
         \label{fig:9qnns50_loss}
     \end{subfigure}
     \hfill
     \begin{subfigure}{0.24\textwidth}
         \centering
         \includegraphics[width=\textwidth]{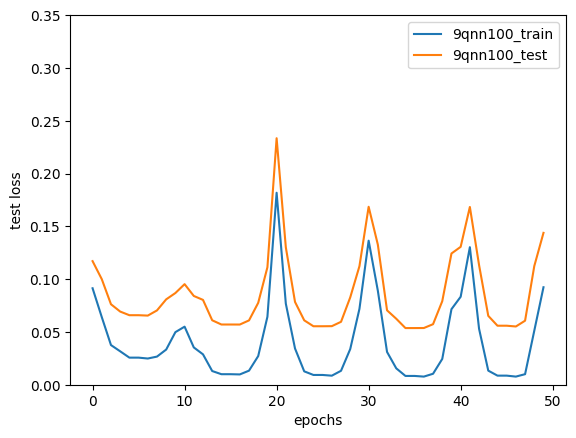}
         \caption{$n_\text{qc}=9$, $d=100$}
         \label{fig:9qnns100_loss}
     \end{subfigure}
     \hfill
     \begin{subfigure}{0.24\textwidth}
         \centering
         \includegraphics[width=\textwidth]{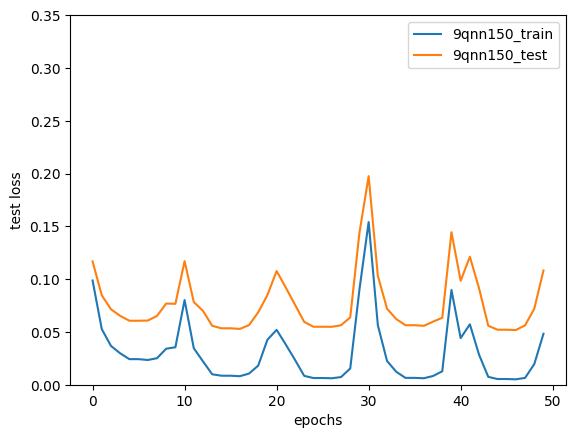}
         \caption{$n_\text{qc}=9$, $d=150$}
         \label{fig:9qnns150_loss}
     \end{subfigure}
     \hfill
     \begin{subfigure}{0.24\textwidth}
         \centering
         \includegraphics[width=\textwidth]{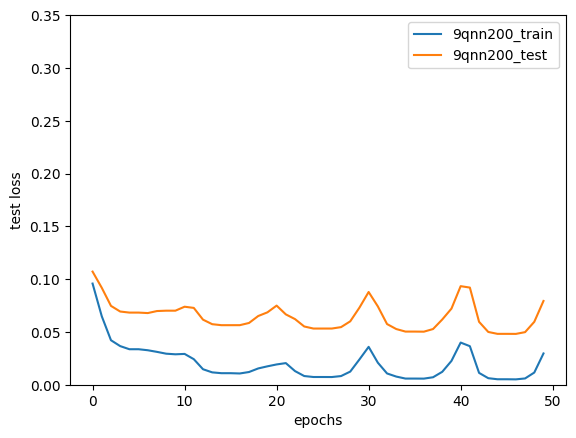}
         \caption{$n_\text{qc}=9$, $d=200$}
         \label{fig:9qnns200_loss}
     \end{subfigure}    
     \\
     \begin{subfigure}{0.24\textwidth}
         \centering
         \includegraphics[width=\textwidth]{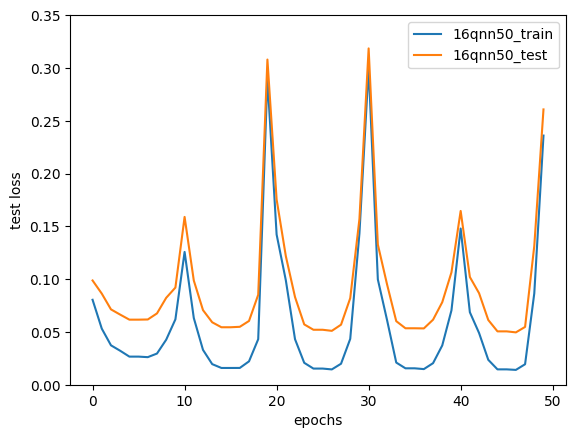}
         \caption{$n_\text{qc}=16$, $d=50$}
         \label{fig:16qnns50_loss}
     \end{subfigure}
     \hfill
     \begin{subfigure}{0.24\textwidth}
         \centering
         \includegraphics[width=\textwidth]{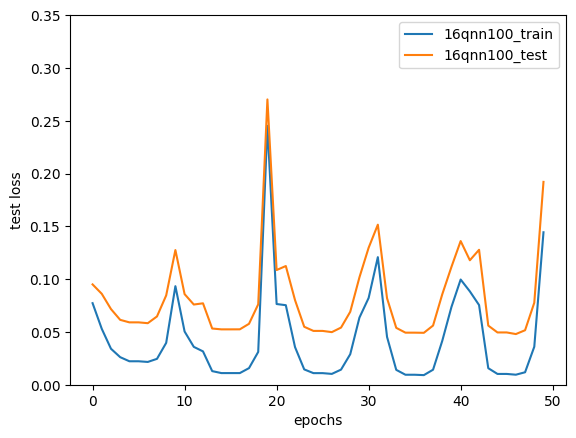}
         \caption{$n_\text{qc}=16$, $d=100$}
         \label{fig:16qnns100_loss}
     \end{subfigure}
     \hfill
     \begin{subfigure}{0.24\textwidth}
         \centering
         \includegraphics[width=\textwidth]{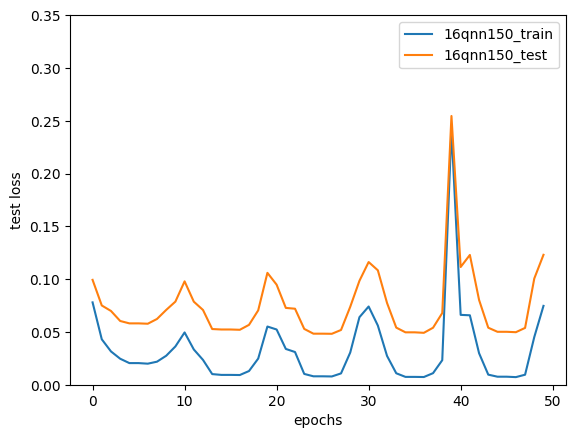}
         \caption{$n_\text{qc}=16$, $d=150$}
         \label{fig:16qnns150_loss}
     \end{subfigure}
     \hfill
     \begin{subfigure}{0.24\textwidth}
         \centering
         \includegraphics[width=\textwidth]{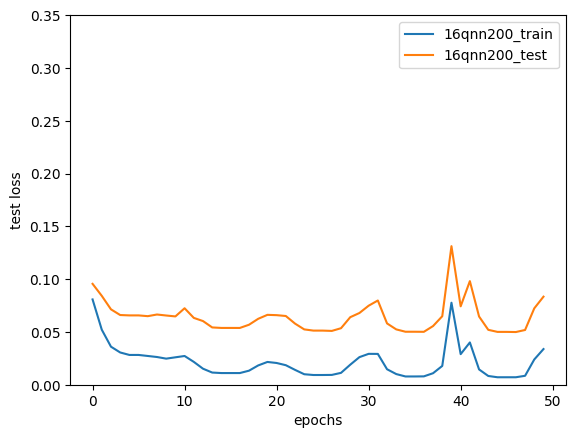}
         \caption{$n_\text{qc}=16$, $d=200$}
         \label{fig:16qnns200_loss}
     \end{subfigure}    
    \caption{
    These figures show the training and test losses at the end of each epoch.
    From these figures, we can see that the losses periodically increase or decrease,  
    because we used Adam with a cosine annealing scheduler in our numerical experiments, as we mentioned in Section~\ref{subsec:settings}. 
    }
    \label{fig:train_test_loss}
\end{figure}




\end{appendices}


\bibliography{sn-bibliography}

@article{cerezo2021variational,
  title={Variational quantum algorithms},
  author={Cerezo, Marco and Arrasmith, Andrew and Babbush, Ryan and Benjamin, Simon C and Endo, Suguru and Fujii, Keisuke and McClean, Jarrod R and Mitarai, Kosuke and Yuan, Xiao and Cincio, Lukasz and others},
  journal={Nature Reviews Physics},
  volume={3},
  number={9},
  pages={625--644},
  year={2021},
  publisher={Nature Publishing Group UK London}
}

@article{peruzzo2014variational,
  title={A variational eigenvalue solver on a photonic quantum processor},
  author={Peruzzo, Alberto and McClean, Jarrod and Shadbolt, Peter and Yung, Man-Hong and Zhou, Xiao-Qi and Love, Peter J and Aspuru-Guzik, Al{\'a}n and O’brien, Jeremy L},
  journal={Nature communications},
  volume={5},
  number={1},
  pages={4213},
  year={2014},
  publisher={Nature Publishing Group UK London}
}

@article{kandala2017hardware,
  title={Hardware-efficient variational quantum eigensolver for small molecules and quantum magnets},
  author={Kandala, Abhinav and Mezzacapo, Antonio and Temme, Kristan and Takita, Maika and Brink, Markus and Chow, Jerry M and Gambetta, Jay M},
  journal={nature},
  volume={549},
  number={7671},
  pages={242--246},
  year={2017},
  publisher={Nature Publishing Group}
}

@article{farhi2014quantum,
  title={A quantum approximate optimization algorithm},
  author={Farhi, Edward and Goldstone, Jeffrey and Gutmann, Sam},
  journal={arXiv preprint arXiv:1411.4028},
  year={2014}
}

@article{farhi2018classification,
  title={Classification with quantum neural networks on near term processors},
  author={Farhi, Edward and Neven, Hartmut},
  journal={arXiv preprint arXiv:1802.06002},
  year={2018}
}

@article{mitarai2018quantum,
  title={Quantum circuit learning},
  author={Mitarai, Kosuke and Negoro, Makoto and Kitagawa, Masahiro and Fujii, Keisuke},
  journal={Physical Review A},
  volume={98},
  number={3},
  pages={032309},
  year={2018},
  publisher={APS}
}

@article{li2018visualizing,
  title={Visualizing the loss landscape of neural nets},
  author={Li, Hao and Xu, Zheng and Taylor, Gavin and Studer, Christoph and Goldstein, Tom},
  journal={Advances in neural information processing systems},
  volume={31},
  year={2018}
}

@article{keskar2016large,
  title={On large-batch training for deep learning: Generalization gap and sharp minima},
  author={Keskar, Nitish Shirish and Mudigere, Dheevatsa and Nocedal, Jorge and Smelyanskiy, Mikhail and Tang, Ping Tak Peter},
  journal={arXiv preprint arXiv:1609.04836},
  year={2016}
}

@article{garipov2018loss,
  title={Loss surfaces, mode connectivity, and fast ensembling of dnns},
  author={Garipov, Timur and Izmailov, Pavel and Podoprikhin, Dmitrii and Vetrov, Dmitry P and Wilson, Andrew G},
  journal={Advances in neural information processing systems},
  volume={31},
  year={2018}
}

@article{fort2019deep,
  title={Deep ensembles: A loss landscape perspective},
  author={Fort, Stanislav and Hu, Huiyi and Lakshminarayanan, Balaji},
  journal={arXiv preprint arXiv:1912.02757},
  year={2019}
}

@article{zhao2020bridging,
  title={Bridging mode connectivity in loss landscapes and adversarial robustness},
  author={Zhao, Pu and Chen, Pin-Yu and Das, Payel and Ramamurthy, Karthikeyan Natesan and Lin, Xue},
  journal={arXiv preprint arXiv:2005.00060},
  year={2020}
}

@article{rudolph2021orqviz,
  title={Orqviz: Visualizing high-dimensional landscapes in variational quantum algorithms},
  author={Rudolph, Manuel S and Sim, Sukin and Raza, Asad and Stechly, Michal and McClean, Jarrod R and Anschuetz, Eric R and Serrano, Luis and Perdomo-Ortiz, Alejandro},
  journal={arXiv preprint arXiv:2111.04695},
  year={2021}
}

@article{huembeli2021characterizing,
  title={Characterizing the loss landscape of variational quantum circuits},
  author={Huembeli, Patrick and Dauphin, Alexandre},
  journal={Quantum Science and Technology},
  volume={6},
  number={2},
  pages={025011},
  year={2021},
  publisher={IOP Publishing}
}

@article{abbas2021power,
  title={The power of quantum neural networks},
  author={Abbas, Amira and Sutter, David and Zoufal, Christa and Lucchi, Aur{\'e}lien and Figalli, Alessio and Woerner, Stefan},
  journal={Nature Computational Science},
  volume={1},
  number={6},
  pages={403--409},
  year={2021},
  publisher={Nature Publishing Group US New York}
}

@article{holmes2022connecting,
  title={Connecting ansatz expressibility to gradient magnitudes and barren plateaus},
  author={Holmes, Zo{\"e} and Sharma, Kunal and Cerezo, Marco and Coles, Patrick J},
  journal={PRX quantum},
  volume={3},
  number={1},
  pages={010313},
  year={2022},
  publisher={APS}
}

@article{mcclean2018barren,
  title={Barren plateaus in quantum neural network training landscapes},
  author={McClean, Jarrod R and Boixo, Sergio and Smelyanskiy, Vadim N and Babbush, Ryan and Neven, Hartmut},
  journal={Nature communications},
  volume={9},
  number={1},
  pages={4812},
  year={2018},
  publisher={Nature Publishing Group UK London}
}

@article{wang2021noise,
  title={Noise-induced barren plateaus in variational quantum algorithms},
  author={Wang, Samson and Fontana, Enrico and Cerezo, Marco and Sharma, Kunal and Sone, Akira and Cincio, Lukasz and Coles, Patrick J},
  journal={Nature communications},
  volume={12},
  number={1},
  pages={6961},
  year={2021},
  publisher={Nature Publishing Group UK London}
}

@article{cerezo2021cost,
  title={Cost function dependent barren plateaus in shallow parametrized quantum circuits},
  author={Cerezo, Marco and Sone, Akira and Volkoff, Tyler and Cincio, Lukasz and Coles, Patrick J},
  journal={Nature communications},
  volume={12},
  number={1},
  pages={1791},
  year={2021},
  publisher={Nature Publishing Group UK London}
}

@article{thanasilp2024exponential,
  title={Exponential concentration in quantum kernel methods},
  author={Thanasilp, Supanut and Wang, Samson and Cerezo, Marco and Holmes, Zo{\"e}},
  journal={Nature communications},
  volume={15},
  number={1},
  pages={5200},
  year={2024},
  publisher={Nature Publishing Group UK London}
}

@inproceedings{you2021exponentially,
  title={Exponentially many local minima in quantum neural networks},
  author={You, Xuchen and Wu, Xiaodi},
  booktitle={International Conference on Machine Learning},
  pages={12144--12155},
  year={2021},
  organization={PMLR}
}

@article{bittel2021training,
  title={Training variational quantum algorithms is NP-hard},
  author={Bittel, Lennart and Kliesch, Martin},
  journal={Physical review letters},
  volume={127},
  number={12},
  pages={120502},
  year={2021},
  publisher={APS}
}

@article{bravyi2016trading,
  title={Trading classical and quantum computational resources},
  author={Bravyi, Sergey and Smith, Graeme and Smolin, John A},
  journal={Physical Review X},
  volume={6},
  number={2},
  pages={021043},
  year={2016},
  publisher={APS}
}

@article{peng2020simulating,
  title={Simulating large quantum circuits on a small quantum computer},
  author={Peng, Tianyi and Harrow, Aram W and Ozols, Maris and Wu, Xiaodi},
  journal={Physical review letters},
  volume={125},
  number={15},
  pages={150504},
  year={2020},
  publisher={APS}
}

@article{pira2023invitation,
  title={An invitation to distributed quantum neural networks},
  author={Pira, Lirand{\"e} and Ferrie, Chris},
  journal={Quantum Machine Intelligence},
  volume={5},
  number={2},
  pages={23},
  year={2023},
  publisher={Springer}
}

@article{marshall2023high,
  title={High dimensional quantum machine learning with small quantum computers},
  author={Marshall, Simon C and Gyurik, Casper and Dunjko, Vedran},
  journal={Quantum},
  volume={7},
  pages={1078},
  year={2023},
  publisher={Verein zur F{\"o}rderung des Open Access Publizierens in den Quantenwissenschaften}
}

@article{kawase2024distributed,
  title={Distributed quantum neural networks via partitioned features encoding},
  author={Kawase, Yoshiaki},
  journal={Quantum Machine Intelligence},
  volume={6},
  number={1},
  pages={15},
  year={2024},
  publisher={Springer}
}

@article{kingma2014adam,
  title={Adam: A method for stochastic optimization},
  author={Kingma, Diederik P},
  journal={arXiv preprint arXiv:1412.6980},
  year={2014}
}

@inproceedings{yao2020pyhessian,
  title={Pyhessian: Neural networks through the lens of the hessian},
  author={Yao, Zhewei and Gholami, Amir and Keutzer, Kurt and Mahoney, Michael W},
  booktitle={2020 IEEE international conference on big data (Big data)},
  pages={581--590},
  year={2020},
  organization={IEEE}
}

@misc{torchquantum2024,
  title={TorchQuantum},
  url={https://github.com/mit-han-lab/torchquantum},
  year={2024}
}

@article{dufter2022position,
  title={Position information in transformers: An overview},
  author={Dufter, Philipp and Schmitt, Martin and Sch{\"u}tze, Hinrich},
  journal={Computational Linguistics},
  volume={48},
  number={3},
  pages={733--763},
  year={2022},
  publisher={MIT Press One Broadway, 12th Floor, Cambridge, Massachusetts 02142, USA~…}
}

@article{kawase2026fast,
  title={Fast and memory-efficient classical simulation of quantum machine learning via forward and backward gate fusion},
  author={Kawase, Yoshiaki},
  journal={arXiv preprint arXiv:2603.02804},
  year={2026}
}

@article{papyan2018full,
  title={The full spectrum of deepnet hessians at scale: Dynamics with sgd training and sample size},
  author={Papyan, Vardan},
  journal={arXiv preprint arXiv:1811.07062},
  year={2018}
}

@article{sagun2016eigenvalues,
  title={Eigenvalues of the hessian in deep learning: Singularity and beyond},
  author={Sagun, Levent and Bottou, Leon and LeCun, Yann},
  journal={arXiv preprint arXiv:1611.07476},
  year={2016}
}

@article{perez2020data,
  title={Data re-uploading for a universal quantum classifier},
  author={P{\'e}rez-Salinas, Adri{\'a}n and Cervera-Lierta, Alba and Gil-Fuster, Elies and Latorre, Jos{\'e} I},
  journal={Quantum},
  volume={4},
  pages={226},
  year={2020},
  publisher={Verein zur F{\"o}rderung des Open Access Publizierens in den Quantenwissenschaften}
}

@article{jerbi2023quantum,
  title={Quantum machine learning beyond kernel methods},
  author={Jerbi, Sofiene and Fiderer, Lukas J and Poulsen Nautrup, Hendrik and K{\"u}bler, Jonas M and Briegel, Hans J and Dunjko, Vedran},
  journal={Nature Communications},
  volume={14},
  number={1},
  pages={517},
  year={2023},
  publisher={Nature Publishing Group UK London}
}

@article{schuld2021effect,
  title={Effect of data encoding on the expressive power of variational quantum-machine-learning models},
  author={Schuld, Maria and Sweke, Ryan and Meyer, Johannes Jakob},
  journal={Physical Review A},
  volume={103},
  number={3},
  pages={032430},
  year={2021},
  publisher={APS}
}

\end{document}